\documentclass[aps,prd,reprint,nofootinbib, preprintnumbers, superscriptaddress]{revtex4-1}

\usepackage{bm}
\usepackage{graphicx} %
\usepackage{url}
\usepackage{subfigure} %
\usepackage{float} %
\usepackage{upgreek} %
\usepackage{multirow} %
\usepackage{color} %
\usepackage{lineno} %
\usepackage{booktabs} %
\usepackage{placeins} %
\usepackage{ulem} %
\usepackage{amsmath} %
\usepackage{amssymb} %
\usepackage{mathrsfs} %
\usepackage{hyperref} %

\usepackage[usenames,dvipsnames]{xcolor}
\hypersetup{colorlinks=true, urlcolor=black, linkcolor=blue, citecolor=red} %
\usepackage{tcolorbox}
\allowdisplaybreaks[4]

\begin{document}
\begin{flushright}
    YITP-23-120
\end{flushright}
\title{One-loop thermal radiation exchange in gravitational wave power spectrum}

\author{Atsuhisa Ota}
\email{aota@cqu.edu.cn}
\affiliation {Department of Physics, Chongqing Key Laboratory for Strongly Coupled Physics, Chongqing University, Chongqing 401331, P.R. China}

\author{Misao Sasaki}
\affiliation{Kavli Institute for the Physics and Mathematics of the Universe (WPI), University of Tokyo, Chiba, 277-8583, Japan}
\affiliation{Center for Gravitational Physics and Quantum Information, Yukawa Institute for Theoretical Physics, Kyoto University, Kyoto, 606-8502, Japan}
\affiliation{Leung Center for Cosmology and Particle Astrophysics, National Taiwan University, Taipei 10617}
\author{Yi Wang}
\affiliation{Department of Physics, The Hong Kong University of Science and Technology, Clear Water Bay, Hong Kong, P.R.China }
\affiliation {HKUST Jockey Club Institute for Advanced Study, The Hong Kong University of Science and Technology, Clear Water Bay, Hong Kong, P.R.China}

\date{\today}

\begin{abstract}

The radiation-dominated universe is a key ingredient of the standard Big Bang cosmology.
Radiation comprises numerous quantum elementary particles, and the macroscopic behavior of radiation is described by taking the quantum thermal average of its constituents. While the interactions between individual particles and gravitational waves are often neglected in this context, it raises the question of whether these elementary particles interact with gravitational waves in the framework of quantum field theory. To address this question, this paper aims to explore the quantum mechanical aspects of gravitational waves in a universe dominated by a massless scalar field, whose averaged energy-momentum tensor plays the role of background radiation. 
 We establish the equivalence between the classical Einstein equation and the mean-field approximation of the Heisenberg equation in a local thermal state. Beyond the mean-field approximation, we analyze the quantum corrections to gravitational waves, particularly focusing on the thermal radiation loop corrections. Interestingly, we find the 1-loop correction surpasses the tree-level spectrum of primordial gravitational waves, which is $O(\alpha^2)$ where $\alpha=H_{\rm inf.}/M_{\rm pl}$ is the ratio of the inflationary Hubble parameter to the Planck mass. Then, to see if this result persists even if we take into account all the higher order loop corrections, the loop expansion is reorganized in the series expansion in $\alpha$. 
 We schematically discuss two-loop diagrams that may give $O(\alpha^2)$ contributions. We leave explicit computations of these diagrams for future studies. Thus, although we cannot claim that the whole loop corrections exceed the tree-level spectrum at the moment, our findings highlight the significance of considering quantum effects when studying the interaction between radiation and gravitational waves in the cosmological context.

%
%
%
%
%
%
% \keywords{Keywords}
%\pacs{04.80.Cc, 95.30.Sf, 98.70.Vc, 98.80.Es}
%
%
\end{abstract}

\maketitle

\section{Introduction}

Our universe was dominated by radiation~(see, e.g.,~\cite{Mukhanov:2005sc}), which consists of numerous elementary particles. Although these particles exhibit quantum behavior, the macroscopic behavior of radiation is described by taking the quantum thermal average of its constituents. By placing the energy-momentum of the fluid on the right-hand side of the Einstein equation, we can solve for the evolution of the universe.
Usually, one derives the equation of motion for gravitational waves in the radiation fluid by linearizing the Einstein equation with the transverse-traceless condition. In this equation, radiation is the driving force behind the expanding universe, while interactions between elementary particles and gravitational waves are neglected. This leads to a natural question: Do these elementary particles interact with gravitational waves at the quantum level?

At first glance, gravitational waves appear minimally coupled to matter contents in general relativity, making them optically thin. However, the universe is comprised of an enormous number of particles. Although the probability of a single interaction process may be negligibly small over cosmic time, quantum features averaged over the universe may be significant.

In the full quantum field theory framework, the classical Einstein equation in a radiation-dominated Universe can be seen as an effective equation of motion resulting from the mean-field approximation of the Heisenberg equation in a local thermal state. In quantum field theory, this approximation is inadequate for obtaining the correct effective action since the interactions at different places are ignored, so a consistent loop analysis is necessary to capture the full quantum effects.

The authors recently discussed similar loop effects during inflation in Refs.~\cite{Ota:2022hvh, Ota:2022xni}.
We evaluated the 1-loop correction for the primordial tensor power spectrum during inflation by considering an enhanced spectator scalar field using the in-in formalism.
In that work, we considered a toy model of an enhanced scalar perturbation and showed that the scalar loops may change the IR amplitude drastically~(see also Refs.~\cite{Inomata:2022yte,Kristiano:2022maq} for recent studies about inflationary loop corrections).
This motivates us to study if a similar effect may exist or not in the radiation-dominated Universe.

In this paper, we consider a similar setup as a first step.  
We consider a massless scalar field in thermal equilibrium and see its quantum effects on the gravitational wave power spectrum.
We emphasize that we take a fully quantum mechanical approach to studying gravitational waves in a universe dominated by thermal radiation, where the quantum averaged energy-momentum tensor is identified as the background.
We demonstrate that the mean-field approximation of the Heisenberg equation reproduces the background Einstein equation and the linear tensor perturbation equation, namely the Einstein equation for gravitational waves. We then go beyond the mean-field approximation and discuss quantum corrections to gravitational waves. We find that the 1-loop spectrum exceeds the tree-level spectrum of primordial gravitational waves. Our result indicates that properly understanding the thermal radiation loop corrections to gravitational waves may be crucial in correctly interpreting the observational data, for example, of the cosmic microwave background~\cite{Planck:2018jri}.

The organization of this paper is as follows.
In Sec.~\ref{sec:eineq}, we derive the linearized Einstein equation for gravitational waves, with a specific focus on the radiation fluid composed of elementary particles. In Sec.~\ref{sec:action}, we reproduce the same result using the action formalism. 
We first derive the equation of motion for gravitational waves without taking the thermal average of the scalar field. At this stage, there appears an effective mass term. However, applying the mean-field approximation, we demonstrate how the effective mass term is eliminated, leading to the standard massless equation of motion. 
In Secs.~\ref{sec:eineq} and~\ref{sec:action}, we review linear cosmological perturbation theory from the equations of motion and action principle points of view, respectively. 
In Sec.~\ref{sec:beyond}, we delve into quantum calculations of the gravitational wave spectrum.
We show that the local vertex diagrams in the 1-loop calculation correspond to the mean-field approximation at the equation of motion level. 
We then evaluate nonlocal vertex diagrams to explore the effects beyond the mean-field approximation, where the 1-loop correction is found to exceed the tree-level spectrum.
In Sec.~\ref{sec:higher}, two-loop order diagrams whose contributions may be as large as or exceed the tree-level spectrum are discussed.
Finally, in Sec.~\ref{sec:conc}, we summarize our results and discuss their physical implications.
Technical details are summarized in the appendices.

\section{Linearized Einstein equation}\label{sec:eineq}

In this section, we derive the linearized Einstein equation for gravitational waves in a radiation-dominated universe, specifically focusing on the radiation constituents.

Let us start with a microscopic description of radiation fluid based on thermal quantum field theory.
For simplicity, we consider a radiation fluid realized by a free massless scalar field $\chi$, but the extension to general gauge fields or fermions is straightforward.
The action of the microscopic system is
\begin{align}
	S_\chi = -\frac{1}{2}\int d^4x  \sqrt{-g}g^{\mu\nu}\partial_\mu \chi \partial_\nu \chi, \label{action1}
\end{align}
where $g_{\mu\nu}$ is the space time metric, and $g$ is the determinant of it.
The energy-momentum tensor of the microscopic massless free scalar field is obtained as
\begin{align}
	T_{\chi,\mu\nu} 
= \partial_\mu \chi \partial_\nu \chi -\frac{g_{\mu\nu}}{2}  g^{\rho\sigma}\partial_\rho \chi \partial_\sigma \chi.\label{eq6} 
\end{align}
Radiation particles are described by quantizing $\chi$, and the fluid dynamical description is realized by taking the thermal average.
After quantization of $\chi$, the microscopic energy-momentum tensor is related to the background radiation energy momentum tensor via 
\begin{align}
	T_{\gamma,\mu\nu} = {\rm Tr}[\hat \varrho_{\chi} \hat T_{\chi,\mu\nu}],\label{tgamma}
\end{align}
where $\hat \varrho_{\chi}$ is a density operator.
Evaluation of Eq.~\eqref{tgamma} in a perturbed spacetime is complicated since the gravitational coupling introduces nonlinear interactions.
However, regardless of microscopic interactions, we consider the local equilibrium state is realized as far as the thermalization time scale is fast enough compared to that of the metric fluctuations~(see Appendixes~\ref{qffflrw} and~\ref{appthermal}). 
Then, the energy-momentum tensor of the locally thermalized radiation fluid should be written in the perfect fluid form
\begin{align}
	T_{\gamma,\mu\nu} = (\rho_\gamma + P_\gamma) u_{\mu}u_{\nu} + P_\gamma g_{\mu\nu},\label{gammaemt}
\end{align}
where $\rho_\gamma$ and $P_\gamma$ are the radiation energy density and pressure measured in the rest frame of the fluid.
We usually describe the radiation dominant Universe after defining the radiation fluid by Eq.~\eqref	{gammaemt}.
Hence, Eq.~\eqref{tgamma} is implicit.
The full action is composed of Eq.~\eqref{action1} and the Einstein-Hilbert action
\begin{align}
	S_{\rm EH} = \frac{M_{\rm pl}^2}{2} \int d^4x  \sqrt{-g} R,\label{action}
\end{align}
where $M_{\rm pl}$ is the reduced Planck mass and $R$ is the Ricci curvature of the spacetime.
Eqs.~\eqref{eq6}, ~\eqref{gammaemt} and the radiation equation of state $P_\gamma=\rho_\gamma/3$ imply that 
\begin{align}
	g^{\mu\nu}T_{\gamma,\mu\nu} = - {\rm Tr}[\hat \varrho_\chi g^{\mu\nu}\partial_\mu \hat \chi \partial_\nu \hat \chi] =0.\label{eq:localdeath}
\end{align}
Hence we find the effective action in the local thermal equilibrium 
\begin{align}
	S = {\rm Tr}[\hat \varrho_\chi (S_\chi + S_{\rm EH})] = S_{\rm EH}. \label{appmean-field}
\end{align}
The Einstein tensor is defined as
\begin{align}
	G_{\mu\nu}\equiv \frac{2}{\sqrt{-g}M_{\rm pl}^2}\frac{\delta S_{\rm EH}}{\delta g^{\mu\nu}},
\end{align}
and then the Einstein equation is found as 
\begin{align}
	G_{\mu\nu}  = \frac{T_{\gamma,\mu \nu} }{ M_{\rm pl}^2}. \label{einstein:eq}
\end{align}
We often consider the linearized Einstein equation for tensor fluctuation defined as
\begin{align}
	g_{ij} = a^2(\delta_{ij}+ \bar h_{ij}),~g_{0i}=0,~g_{00}=-a^2,
\end{align}
where $a$ is the isotropic background scale factor.
$\bar h_{ij}$ is transverse-traceless with respect to the background spatial metric: $\delta^{ij}\partial_i \bar h_{jk} = \delta^{ij}\bar h_{ij}=0$. 
Linearizing the Einstein equation $G_{ij} = T_{\gamma,ij}/{M_{\rm pl}^2}$, we find
\begin{align}
	\bar  h''_{ij} + 2\frac{a'}{a} \bar h_{ij}' -\partial^2 \bar h_{ij} + m_{\rm eff}^2\bar h_{ij} = 0\label{lineqij},
	\\
	m_{\rm eff}^2 \equiv  2 \left(\frac{a'^2}{a^2} -  2\frac{a''}{a}\right) -\frac{2}{M_{\rm pl}^2}a^2 P_\gamma.
\end{align}
Using the Friedmann equations
\begin{align}
	\frac{a''}{a} &= \frac{a^2(\rho_\gamma - 3P_\gamma)}{6M_{\rm pl}^2},
	\\
	\frac{a'^2}{a^2} &= \frac{a^2 \rho_\gamma }{3M_{\rm pl}^2},
\end{align} 
we find the equation of motion for the massless gravitational waves
\begin{align}
	\bar h''_{ij} + 2\frac{a'}{a} \bar h'_{ij} -\partial^2 \bar h_{ij} =0.\label{eomtens}
\end{align}

When we establish Einstein equation~\eqref{einstein:eq}, the interaction of gravitational waves and microscopic radiation fields is integrated out by Eq.~\eqref{eq:localdeath}.
As shown in Appendix~\ref{appthermal}, Eq.~\eqref{appmean-field} is evaluated in a local thermal state by using the free radiation field, which corresponds to the \textit{mean-field approximation} in quantum field theory in the sense that we ignored the iterative solutions with respect to cosmological perturbations.
The vanishing local interaction implies that the equivalence principle locally eliminates gravitational waves.
However, as we will see later in this paper, the coordinate transformation cannot simultaneously vanish gravitational waves at different locations, which is not captured in the mean-field approximation.
In this paper, we evaluate the gravitational wave spectrum and investigate the validity of Eq.~\eqref{eomtens} in the presence of quantum corrections.

\section{Action formalism}\label{sec:action}

Before going beyond the mean-field approximation, let us find the same result~\eqref{eomtens} based on the action principle for cosmological perturbations.
While we reproduce the same result, revisiting the action formalism is useful to see the gap between the fluid dynamical approximation and the full quantum approach in the next section. As a by-product, we will also see that different definitions of second-order tensor modes eventually give consistent results.

We expand the action to the second order in cosmological perturbations and find the equation of motion.
Then, we take the thermal average over the radiation field.
Let us start with a metric in the 3+1 form:
\begin{align}
	ds^2 = -N^2 d\tau^2 + \gamma_{ij}(N^i d\tau +dx^i)(N^j d\tau +dx^j).
\end{align}
There is an ambiguity in the definition of tensor fluctuations at second order.
Here, we perturb the spatial metric with perturbation variables on the exponent:
\begin{align}
	\gamma_{ij} &= a^2 \delta_{ik}(e^h)^k{}_j 
	\notag \\
	&= a^2 \delta_{ik}\left( \delta^k{}_j + h^k{}_j + \frac12 h^k{}_l h^l{}_j +\cdots \right),\label{eqexph}
	\\
	\gamma^{ij} &= a^{-2} (e^{-h})^i{}_k\delta^{kj} \notag \\
	&
	= a^{-2} \left( \delta^i{}_k - h^i{}_k + \frac12 h^i{}_l h^l{}_k +\cdots \right)\delta^{kj},
\end{align}
with the transverse-traceless condition:
\begin{align}
	\partial_i h^i{}_j = h^i{}_i = 0.
\end{align}
For simplicity, we ignore first-order scalar and vector perturbations.
We consider $h^i{}_j$ generated during inflation, whose typical amplitude is the Hubble-to-Planck mass ratio.
We may not take the lapse and shift freely at second order even if we ignore pure second-order scalar and vector perturbations since tensor perturbations introduce nonvanishing scalar and vector modes in the second order.
However, the background equation of motion satisfies the Hamiltonian and momentum constraints so that the second-order lapse and shift are eliminated from the second-order action.
In the exponentiated metric, the volume element $\sqrt{-g}$ is unperturbed by $h^i{}_j$, simplifying the perturbative expansion.
The full action is composed of Eqs.~\eqref{action1} and \eqref{action}:
\begin{align}
	S = S_\chi + S_{\rm EH}.
\end{align}
In the previous section, we directly integrated out the scalar field in Eq.~\eqref{eq:localdeath}, but we keep the scalar field in this section to see how the approximation works to cancel the mass at linear order.
The action of the minimally coupled scalar radiation is expanded into 
\begin{align}
	S_\chi = S_{\chi0} + S_{h\chi \chi}+ S_{hh \chi\chi}+\cdots,
\end{align}
where we defined 
\begin{align}
	S_{\chi0} &= \frac{1}{2}\int d^4 x ~ a^2 (\chi'^2  - \delta^{ij}\partial_i \chi \partial_j \chi),\label{chi0action}
	\\
	S_{h\chi \chi} &= \frac{1}{2}  \int d^4 x ~ a^2 h^i{}_k \delta^{kj}\partial_j \chi \partial_i \chi ,\label{hchichi}
	\\
	S_{hh \chi\chi} & = -\frac{1}{4}  \int d^4 x ~ a^2 h^i{}_k h^k{}_l \delta^{lj}\partial_j \chi \partial_i \chi.\label{hhchichi}
\end{align}
The Einstein-Hilbert action is expanded into
\begin{align}
	S_{\rm EH} = S_{\rm bg} + S_{h0},
\end{align}
where $S_{\rm bg} $ is a functional of the scale factor, and we defined
\begin{align}
	S_{\rm bg} &= -3 M_{\rm pl}^2 \int d^4 x ~ a'^2 \label{Sbg},
	\\
	S_{h0} &= \frac{M_{\rm pl}^2}{8}\int d^4 x ~ a^2\left( h'^i{}_jh'^j{}_i  - \delta^{kl} \partial_k h^i{}_j \partial_l h^j{}_i\right).\label{action:h0}
\end{align}
Variation of the full action with respect to the metric perturbation is then
\begin{align}
	\frac{\delta S}{\delta h^{i}{}_{j}} &= -\frac{M_{\rm pl}^2}{4}a^2 \left( h''^j{}_i + 2\frac{a'}{a} h'^j{}_i - \partial^2 h^j{}_i \right)
	\notag 
	\\
	&+ \frac{1}{2}a^2 \delta^{jl}\partial_l  \chi \partial_i  \chi
	- \frac{1}{2}a^2 h^{j}{}_k \delta^{kl}\partial_l  \chi \partial_i  \chi.
\end{align}
Also, the chain rule with respect to the metric implies 
\begin{align}
	\frac{\delta S}{\delta h^{i}{}_{j}} = \frac{\delta S}{\delta g^{\mu\nu}} \frac{\delta g^{\mu\nu}}{\delta h^{i}{}_{j}} 
	=0,
\end{align}
where the last equality follows from the variation of the whole system with respect to $g^{\mu\nu}$. i.e., $\delta S/\delta g^{\mu\nu}=0$.
To summarize, we find
\begin{align}
	  &h''^j{}_i + 2\frac{a'}{a} h'^j{}_i - \partial^2 h^j{}_i + \frac{2}{M_{\rm pl}^2} h^{j}{}_k \delta^{kl}\partial_l  \chi \partial_i  \chi 
	  \notag 
	  \\
	  &=
	\frac{2}{M_{\rm pl}^2} \delta^{jl}\partial_l  \chi \partial_i  \chi  . \label{eomtens:2}
\end{align}
The effective mass and source appear at the microscopic level.
By taking the thermal average, this equation of motion reduces to Eq.~\eqref{eomtens} as follows.
First, Eqs.~\eqref{eq6},~\eqref{tgamma} and~\eqref{eq:localdeath} yield
\begin{align}
	{\rm Tr}[\hat \varrho_{\chi} \partial_\mu \hat \chi \partial_\nu \hat \chi] = (\rho_\gamma + P_\gamma) u_{\mu}u_{\nu} + P_\gamma g_{\mu\nu} .
\end{align}
We only consider tensor metric perturbations in an FLRW background, so we have $u_i = 0$.
Hence, we find 
\begin{align}
	{\rm Tr}[\hat \varrho_{\chi} \partial_i \hat \chi \partial_j \hat \chi] = P_\gamma a^2 ( \delta_{ij}  +  h_{ij} +\cdots), \label{chiichij}
\end{align}
where the dots imply the higher order term in the tensor fluctuations.
The mass term is canceled by substituting Eq.~\eqref{chiichij} into Eq.~\eqref{eomtens:2}.
Also, the source reduces to the trace component, which vanishes at the action level.
Thus we reproduce Eq.~\eqref{eomtens}.

As we mentioned, there is an ambiguity about the definition of tensor perturbations in the second order.
For example, instead of Eq.~\eqref{eqexph}, we may consider 
\begin{align}
	\gamma_{ij} &= a^2 (\delta_{ij} + \bar h_{ij}),
	\\
	\gamma^{ij} &= a^{-2} (\delta^{ij} - \bar h^{ij} - \bar h^{ik}\bar h_k{}^{j}+\cdots ),
\end{align}
with the transverse-traceless condition
\begin{align}
	\delta^{ik}\partial_k \bar h_{ij} = \delta^{ij} \bar h_{ij} = 0.
\end{align}
This convention has a different second-order action, but the linearized Einstein equation should be the same. 
All terms related to gravitational waves are expanded into
\begin{align}
	S_{\bar h0} =& \frac{M_{\rm pl}^2}{8}\int d^4 x ~ a^2 \left[ \delta^{ik}\delta^{jl}\bar h'_{ij}\bar h'_{kl} - \delta^{mn} \delta^{ik}\delta^{jl} \partial_m \bar h_{ij} \partial_n \bar h_{kl} 
	\right] 
	\notag \\
	&+ \frac{M_{\rm pl}^2}{2}\int d^4 x\left(\frac{1}{2}  a'^2  - a a''   \right) \delta^{ik}\delta^{lj}\bar h_{ij}\bar h_{kl},
\end{align}
where the second line is the perturbation of the volume element, which is absent for the gauge choice~\eqref{eqexph}.
The radiation field action is also expanded into
\begin{align}
	S_{\bar h\chi \chi}&=
	 	\frac{1}{2}  \int d^4 x ~ a^2  \delta^{ik}\delta^{jl}\bar h_{ij} \partial_k \chi \partial_l \chi,
	\\
	S_{\bar h\bar h\chi \chi}& = -\frac{1}{2}  \int d^4 x ~ a^2 \delta^{in}\delta^{kj}\delta^{lm} \bar h_{ij}\bar h_{kl}\partial_m \chi \partial_n \chi 
	\notag 
	\\
	&-\frac{1}{8}  \int d^4 x ~ a^2
	  \delta^{ik}\delta^{jl}\bar h_{ij}\bar h_{kl} \left(\chi'^2
	  -
	  \delta^{mn}\partial_m \chi\partial_n \chi
 \right)
	 .	
\end{align}
Variation of the full action, $\delta S/\delta \bar h^{ij}=0$ is
\begin{align}
 & \bar h''_{ij} + 2\frac{a'}{a} \bar  h'_{ij} - \partial^2 \bar h_{ij}		+ \frac{2}{M_{\rm pl}^2} \bar h_{ik}\delta^{kl}\partial_l  \chi \partial_j  \chi
	\notag 
	\\
	&+\frac{2}{M_{\rm pl}^2} \bar h_{kj}\delta^{kl}\partial_i  \chi \partial_l  \chi
 -  \mu^2 \bar h_{ij} = 
	\frac{2}{M_{\rm pl}^2} \partial_i  \chi \partial_j  \chi,
 \label{eomhbar}
\end{align}
where we find
\begin{align}
	 \mu^2 = 2  \frac{a'^2}{a^2}  -  4 \frac{a''}{a}   
 -\frac{1}{M_{\rm pl}^2}
	  \left( \chi'^2
	  -
	  \delta^{mn}\partial_m  \chi\partial_n  \chi
 \right).
\end{align}
Taking the thermal average by using Eq.~\eqref{chiichij}, we reproduce Eq.~\eqref{eomtens} again.
Thus, the ambiguity of field redefinition in the 4-point interaction is eliminated.

\begin{figure*}
\includegraphics[width=\linewidth]{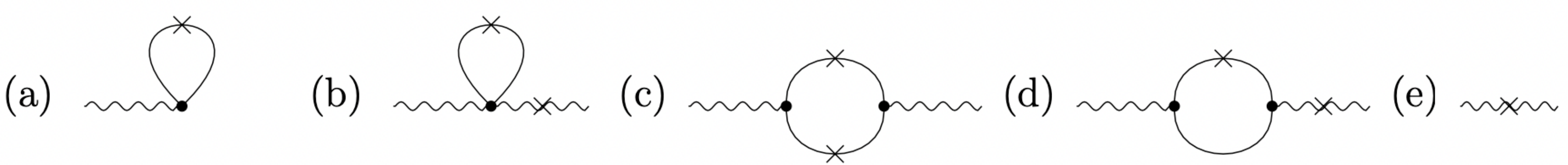}
\caption{Diagrammatic representation of 1-loop gravitational wave spectrum.
Propagators with and without a cross imply the Keldysh and retarded Green functions defined in Appendix~\ref{app:green}.
Solid and wavy lines represent the scalar field and gravitational waves, respectively. Diagrams (a) and (b) can be derived from the mean-field approximation in Sections~\ref{sec:eineq} and \ref{sec:action}. However, the tadpole (a) in a local thermal state introduces an additional external gravitational wave leg, which cancels the local diagram (b). Beyond the mean-field approximation for the equation of motion,
(c) is gravitational waves induced by the inhomogeneous scalar field calculated in Sec~\ref{sec:mf}, which turns out to be suppressed in the IR limit; and (d) denotes the radiation exchange in gravitational waves, calculated in Sec~\ref{sec:fis}, which gives the dominant contribution.
The tree-level power spectrum is drawn as (e).
}
\label{fnpow}	
\end{figure*}

The origin of Eq.~\eqref{chiichij} is essential to find the massless linearized Einstein equation. Fluid dynamically, Eq.~\eqref{chiichij} is given as a generally covariant ansatz of the energy-momentum tensor in the lowest order of the gradient expansion. We can reproduce Eq.~\eqref{chiichij} in thermal field theory in a perturbed FLRW spacetime as discussed in appendix~\ref{appthermal}.
In the appendix, we consider the free field $\chi$ in an FLRW spacetime and then take the trace for a local thermal state density operator to reproduce the same result by using the tetrads.
Hence, the local thermal average of the equation of motions~\eqref{eomtens:2} and \eqref{eomhbar} corresponds to taking the mean-field approximation in a global FLRW frame with a local thermal state density operator.
This contribution is the 1-vertex diagram from the 4-point interaction in a diagrammatic perspective, as shown in (a) and (b) in Fig.~\ref{fnpow}.
The 3-point interaction introduces 2-vertex nonlocal 1-loop diagrams, such as (c) and (d) in the figure, which are not considered in the mean-field approximation.
The vanishing 1-vertex diagram implies that gravitational waves are locally eliminated; however, the same argument does not apply to the nonlocal vertex diagrams since we cannot eliminate gravitational waves in different places simultaneously.

\section{Radiation exchange}\label{sec:beyond}

In this section, we compute the time evolution of the tensor fluctuation in the in-in formalism and go beyond the tree-level approximation.
Given a free action of gravitational wave~\eqref{action:h0} and its interactions with the radiation field~\eqref{hchichi} and~\eqref{hhchichi}, we may consider nonlinear evolution of gravitational waves in the perturbative expansion.
Using the interaction Hamiltonian $\hat H_{\rm int}$, the interaction picture time evolution operator $\hat F$ is~\cite{Weinberg:2005vy}
\begin{align}
	&\hat F(\tau;\tau_0) = 1 - i \lambda \int^\tau_{\tau_0} d\tau' \hat H_{{\rm int}}(\tau')
	\notag 
	\\
	&-\lambda ^2\int^\tau_{\tau_0} d\tau' \int^{\tau'}_{\tau_0} d\tau'' \hat H_{{\rm int}}(\tau')\hat H_{{\rm int}}(\tau'')+\mathcal O(\lambda^3),\label{defFop}
\end{align}
where $\lambda$ is the order counting parameter of the interaction Hamiltonian.
Then the Heisenberg operator $\hat {\mathcal O}_{\rm H}$ evolves as
\begin{align}
	\hat {\mathcal O}_{\rm H}(\tau) =& \hat F(\tau;\tau_0)^\dagger \hat {\mathcal O}_{\rm I}(\tau) \hat F(\tau;\tau_0), \label{Heisen-hop}
\end{align}
where the subscript ``I'' implies the interaction picture field, which hereafter we omit unless otherwise stated.
Eq.~\eqref{Heisen-hop} is expanded into~\cite{Weinberg:2005vy}
\begin{align}
	&\hat {\mathcal O}_{\rm H}(\tau) = \hat {\mathcal O}(\tau) + i \lambda \int^\tau_{\tau_0} d\tau' [\hat H_{{\rm int}}(\tau'),\hat {\mathcal O}(\tau)]
	\notag \\
	&-\lambda ^2\int^\tau_{\tau_0} d\tau' \int^{\tau'}_{\tau_0} d\tau'' [\hat  H_{{\rm int}}(\tau''), [\hat H_{{\rm int}}(\tau'),\hat {\mathcal O}(\tau)]]+\mathcal O(\lambda^3).\label{defOevol}
\end{align}
The interaction picture fields expand interaction Hamiltonians as 
\begin{align}
	\hat H_{\rm int} = \epsilon \hat H_{h\chi\chi} + \epsilon^2 \hat H_{hh\chi\chi} +\mathcal O(\epsilon^3),
\end{align}
where $\epsilon$ counts order in cosmological perturbations and the Legendre transformation of Eqs.~\eqref{hchichi} and \eqref{hhchichi} yields
\begin{align}
	\hat H_{h\chi\chi} &=- \frac{ a^2}{2} \int d^3 x ~ \hat h^{ij}\partial_i \hat \chi \partial_j \hat \chi,\label{Hhchichi}
	\\
	\hat H_{hh\chi\chi} &= \frac{ a^2}{4} \int d^3 x ~ \hat h^{ik} \hat h_k{}^j\partial_i \hat \chi \partial_j \hat \chi \label{Hhhchichi}.
\end{align}
In the following, we evaluate Eq.~\eqref{defOevol} for the tensor fluctuation $\hat {\mathcal O} = \hat h_{ij}$ order by order in $\lambda$ and $\epsilon$.
Then we compute the power spectrum of the Heisenberg operator $\hat h_{{\rm H},ij}$.
One can also directly evaluate the time evolution of $\hat {\mathcal O} =\hat h_{ij}\hat h^{ij}$ in Eq.~\eqref{defOevol}. The equivalence of these approaches is guaranteed since $\hat F$ is unitary.
However, the former approach is more convenient since the physical interpretation of each term is clearer than the latter. 
Several cancellations of divergences may appear in the latter approach, and the physical interpretation of the expansion is not straightforward.  
We summarize the correction to the tensor fluctuation in the following form:
\begin{align}
	\delta \hat h_{ij} =\sum_{n,m} \lambda^n \epsilon^{m}	\delta \hat h^{(n,m)}_{ij},
\end{align}
where we defined the 1-loop order corrections
\begin{align}
	\delta \hat h^{(1,2)}_{ij}(x) &=i \int^{x^0} dy^0 [\hat H_{h\chi\chi}(y^0),\hat h_{ij}(x)],\label{h12}
	\\
	\delta \hat h^{(1,3)}_{ij}(x) &=i \int^{x^0} dy^0 [\hat H_{hh\chi\chi}(y^0),\hat h_{ij}(x)],\label{h13}
	\\
	\delta \hat h^{(2,3)}_{ij}(x) &=i\int^{x^0} dy^0i \int^{y^0} dz^0 
	\notag 
	\\
	&\times 
	\left[ \hat H_{h\chi\chi}(z^0),    \left[\hat H_{h\chi\chi}(y^0),\hat h_{ij}(x)\right]\right]. \label{h23}
\end{align}
We truncate the expansion at this order since higher order corrections in terms of $\lambda$ and $\epsilon$ do not contribute to the 1-loop order power spectrum.
The expectation value of operators is
\begin{align}
	\langle \hat {\mathcal O}_{\rm H} \rangle \equiv  {\rm Tr}\left[\hat \varrho \hat {\mathcal O}_{\rm H}\right],
\end{align}
where $\hat \varrho$ is a density operator that defines a quantum state, which is written as
\begin{align}
	\hat \varrho \simeq  \hat \varrho_{\chi} \otimes \hat \varrho_h.
\end{align}
In our case, $\chi$ is in a local thermal state $\hat \varrho_\chi$ such that Eq.~\eqref	{tgamma} is satisfied.
The trace of $\hat \varrho_{\chi}$ will be defined in a local frame (see Appendix~\ref{appthermal} for details).
A state $\hat \varrho_h$ for gravitational waves is defined as the minimum energy state in the remote past during inflation, which is not a vacuum state at the initial time of radiation dominance (see Appendix~\ref{apppgw} for the definition).

\subsection{$\mathcal O(\lambda)$: mean field in operator evolution}\label{sec:mf}

Let us first consider the $\mathcal O(\lambda)$ contribution, involving only one interaction Hamiltonian in the time evolution of the Heisenberg operator \eqref{defOevol}. In terms of the Heisenberg operator itself, this may be understood as a mean-field approximation~(or Born approximation). However, as we will take a trace of the thermal density matrix, this so-called ``Born approximation'' here can still generate loop diagrams.
The 3-point interaction \eqref{Hhchichi} and Eq.~\eqref{h12} yield
\begin{align}
\delta \hat h^{(1,2)}_{ij}(x)	&=
	\frac{i}{2} \int^{x^0} dy^0 a(y^0)^2 \int d^3 y \notag \\
	&\times [  \hat h_{ij}(x) , \hat h^{kl}(y) ]	\partial_k \hat \chi(y) \partial_l \hat \chi(y).
\end{align}
Using the retarded Green function~\eqref{rt:h}, we find $\mathcal O(\lambda\epsilon^2)$ solution 
\begin{align}
	\delta \hat h^{(1,2)}_{ij}(x)		& =\frac{2}{M_{\rm pl}^2} \int d^4 y ~ G_{ij}{}^{kl}(x,y) \partial_k \hat \chi(y) \partial_l \hat \chi(y).\label{eqh12}
\end{align}
Similarly, the 4-point interaction \eqref{Hhhchichi} and Eq.~\eqref{h13} yield $\mathcal O(\lambda\epsilon^3)$ solution
\begin{align}
	\delta \hat h^{(1,3)}_{ij}(x) &= -\frac{2}{M_{\rm pl}^2}\int d^4 y ~ G_{ij}{}^{kl}(x,y) 
	\notag 
	\\
	&\times \hat h_l{}^m(y)\partial_k \hat \chi(y) \partial_m \hat \chi(y).
\end{align}
To summarize, at $\mathcal O(\lambda)$, we find 
\begin{align}
	&\delta \hat h^{(1,2)}_{ij}(x) + \delta \hat h^{(1,3)}_{ij}(x)
	= \frac{2}{M_{\rm pl}^2}\int d^4 y ~ G_{ij}{}^{kl}(x,y)
	\notag 
	\\
	&\times  (\delta_l{}^m - \hat h_l{}^m(y))\partial_k \hat \chi(y) \partial_m \hat \chi(y).\label{58}
\end{align}

Now, we are ready to compute the 1-loop order power spectrum from $\mathcal O(\lambda)$ field~\eqref{58}.
Eq.~\eqref{58} yields two types of contributions to the 1-loop power spectrum of the gravitational waves.
One is the cross-correlation function of Eq.~\eqref{58} and the linear tensor perturbation, and another is the auto-correlation function of Eq.~\eqref{58}.
These are called cut in the side~(CIS) diagram and cut in the middle~(CIM) diagram, respectively~\cite{Senatore:2009cf}.
The CIS diagram is written as
\begin{align}
{\rm Tr}[\hat \varrho  \hat h^{ij} ( \delta \hat h^{(1,2)}_{ij} + \delta \hat h^{(1,3)}_{ij} ) +{\rm 1~perm.}],
\end{align}
which reduces to 
\begin{align}
{\rm Tr}[\hat \varrho_h  \hat h^{ij} {\rm Tr}[\hat \varrho_\chi ( \delta \hat h^{(1,2)}_{ij} + \delta \hat h^{(1,3)}_{ij} ) ]+{\rm 1~perm.}].
\end{align}
Thus, we take the partial trace with respect to $\chi$, and then take the trace for the tensor mode.
The $\chi$ average is evaluated as
\begin{align}
	&{\rm Tr}[\hat \varrho_\chi  ( \delta \hat h^{(1,2)}_{ij}(x) + \delta \hat h^{(1,3)}_{ij}(x) ) ] 
	\notag 
	\\
	=&
	\frac{2}{M_{\rm pl}^2}\int d^4 y ~ G_{ij}{}^{kl}(x,y)(\delta_l{}^m - \hat h_l{}^m(y))  
	\notag 
	\\
	&\times  {\rm Tr}[\hat \varrho_\chi \partial_k \hat \chi(y) \partial_m \hat \chi(y)] 
	\notag 
	\\
	=&
	\frac{2}{M_{\rm pl}^2}\int d^4 y ~ G_{ij}{}^{kl}(x,y) (\delta_l{}^m - \hat h_l{}^m(y))  
	\notag 
	\\
	&\times   P_\gamma a^2 ( \delta_{km}  +  \hat h_{km}(y)) 
	\notag 
	\\
	=&\mathcal O(\epsilon^2),
\end{align}
where we used Eq.~\eqref{chiichij} in the second line, and $G_{ij}{}^{kl}\delta_{kl}=0$ for the last line.
This cancellation at linear order implies that the effective mass is zero, consistent with the mean-field approximation in the previous sections.
From a diagrammatic perspective in Fig.~\ref{fnpow}, this is the cancelation of the 1-vertex diagram (b) and the 1-loop tadpole graph (a), perturbed by the local thermal state density operator.
Diagram (b) depends on the choice of second-order tensor fluctuation and, thus, the field redefinition.
This cancellation implies that the field redefinition dependence is removed by the tadpole.

The 1-loop CIM diagram is the autocorrelation of Eq.~\eqref{58}, which is thermally induced gravitational waves.
Introducing the scalar Keldysh Green function at finite temperature~\eqref{kd:chi}, the \textit{equal time} 2-point correlation function of the induced gravitational waves is written as

\begin{align}
	&{\rm Tr}[\hat \varrho \delta \hat h^{(1,2)}_{ij}(x)\delta \hat h^{(1,2)ij}(x') ]
	\notag 
	\\
	&=\frac{4}{M_{\rm pl}^4}  \int d^4 y \int d^4 z
	G_{ij}{}^{kl}(x,y) G^{ijmn}(x',z)
	\notag 
	\\
	&\times 
	{\rm Tr} \left[\hat \varrho_\chi \partial_k \chi(y) \partial_l \chi(y)
	   \partial_m \chi(z) \partial_n \chi(z) \right]
    \notag 
	\\
	&=\frac{8}{M_{\rm pl}^4}  \int d^4 y \int d^4 z
	G_{ij}{}^{kl}(x,y) G^{ijmn}(x',z)
	\notag 
	\\
	&\times 
	{\rm Tr} \left[\hat \varrho_\chi \partial_k \chi(y)  \partial_m \chi(z)\right]
	  {\rm Tr} \left[\hat \varrho_\chi  \partial_l \chi(y) \partial_n \chi(z) \right]
   \notag 
	\\
	&=\frac{8}{M_{\rm pl}^4}  \int d^4 y \int d^4 z
	G_{ij}{}^{kl}(x,y) G^{ijmn}(x',z)
	\notag 
	\\
	&\times 
    \partial_k^{(y)} \partial_m^{(z)}G^K(y,z)
    \partial_l^{(y)} \partial_n^{(z)}G^K(y,z).\label{hhind}
\end{align}
where the trispectrum of the scalar fields is reduced to a product of power spectra by using Wick's theorem at finite temperature~\eqref{wick:finite}, and each Keldysh Green function is calculated by Eq.~\eqref{GKexp}. 
We then rewrite the Green functions to simplify this equation using the Fourier transform \eqref{green:tens}.
Finally, we integrate out the real space coordinates and then the momenta.
The angular integral is simplified using Eq.~\eqref{C7} and~(See also an appendix of Ref.~\cite{Ota:2022xni}.)
\begin{align}
	& \int \frac{d^3 k_1 d^3 k_2}{(2\pi)^6}(2\pi)^3\delta(\mathbf k_1+\mathbf k_2 -\mathbf q)f(k_1,k_2,q)
	\notag 
	\\
	& = \frac{1}{(2\pi)^2q} \int_0^\infty dk_2 \int_{|k_2-q|}^{k_2+q} dk_1 k_1 k_2f(k_1,k_2,q).\label{C8}
\end{align}
After simplifying Eq.~\eqref{hhind}, we find the power spectrum in Fourier space
\begin{align}
		&{\rm Tr}\left[ \hat \varrho_\chi \delta \hat h^{(1,2)}_{ij}(x^0,\mathbf q) \delta \hat h^{(1,2)ij}(x^0,\mathbf q')\right] 
	\notag 
	\\
	&=	(2\pi)^3\delta(\mathbf q +\mathbf q') P^\text{1-loop}_{h,{\rm ind.}}(x^0,q),
\end{align}
where we defined
\begin{align}
P^\text{1-loop}_{h,{\rm ind.}}(x^0,q)		
&= \frac{2}{q^2 M_{\rm pl}^4a(x^0)^2}   
\notag 
\\
&\times 
	 \int_0^{\infty} \frac{dp}{p} \int^{|p+q|}_{|p-q|} \frac{dp'}{p'} \bar w(q,p,p')
	 \notag 
	 \\
	 &\times  \mathcal I (x^0,q,p,p')f_{\beta p}f_{\beta p'}.\label{indeq}
\end{align}
Note that we defined the angular factor
\begin{align}
	\bar w(q,p,p') &= \frac{pp'}{128 \pi ^2 q^5}(q-p-p')^2 (q+p-p')^2
	\notag
	\\
	&\times  (q-p+p')^2 (q+p+p')^2,
\end{align}
and the Plank distribution
\begin{align}
	f_{\beta p} \equiv \frac{1}{e^{\beta p}-1}.
\end{align}
Note that $p$ is the comoving momentum, and $\beta$ is the inverse of the comoving temperature, hence constant in time.
The time integral is summarized as
\begin{align}
	\mathcal I(x^0,q,p,p')
	&\equiv \int^{x^0}_0 \frac{dy^0}{a(y^0)}
	\int^{x^0}_0 \frac{dz^0}{a(z^0)}
	\notag 
	\\
	&\times 
	\sin (q(x^0-y^0))
	\sin (q(x^0-z^0))
	\notag 
	\\
	&\times 
	\cos (p(y^0-z^0))
	\cos (p'(y^0-z^0)).
\end{align}
Typical momenta of the radiation field are given as $p\sim p'\sim \beta^{-1}$ from the Planck distribution.
The Friedmann equation gives $1/(a\beta)^{4}\sim M_{\rm pl}^2 H^2$, which yields 
\begin{align}
	\beta \sim \sqrt{\frac{H}{M_{\rm pl}}}\frac{1}{aH} \ll \frac{1}{aH},\label{hightemplim}
\end{align}
where $H$ is the Hubble parameter.
$x^0$, $y^0$ and $z^0$ are the time scale of cosmic expansion so that we may consider $\beta \ll x^0,y^0,z^0$ and $p^{-1},p'^{-1} \ll x^0,y^0,z^0$.
For IR gravitational waves, $q(x^0 - y^0)\ll 1$, one finds
\begin{align}
	\mathcal I (x^0,q,p,p') \to q^2 \tilde {\mathcal I} (x^0,p,p').
\end{align}
The IR asymptotic formula for $P^\text{1-loop}_{h,{\rm ind.}}(x^0,q)$ is found as
\begin{align}
&\lim_{q\to 0}P^\text{1-loop}_{h,{\rm ind.}}(x^0,q)		
\notag 
\\
&= \frac{4}{15 \pi ^2 M_{\rm pl}^4a(x^0)^2}   
	 \int_0^{\infty} dp  p^4
 \tilde{\mathcal I} (x^0,p,p)f_{\beta p}^2
 ,\label{indeqir}
\end{align}
where, for $p\gg q$, we used
\begin{align}
	\int_{p-q}^{p+q} dp' \bar w(q,p,p') F(p,p') 
	=\frac{2 p^6 F(p,p)}{15 \pi ^2} +\mathcal O\left(q^2\right),\label{IRform}
\end{align}
for an arbitrary function $F(p,p')$.
Note the $q$-dependence both in the integral domain and in the integrand of the above, which makes it finite in the limit $q\to0$.  This leads to the $q$-independent result given by Eq.~\eqref{indeqir}, irrespective of the integrand of the remaining $p$ integral.
Thus, in the IR region, we find a white noise spectrum for the induced GWs.
Hence, the dimensionless power spectrum is suppressed on super-horizon scales, as expected from the causal origin of the induced GWs. 
In the high-temperature limit, $\beta\ll x^0$, Eq.~\eqref{IRform} is evaluated as
\begin{align}
 &\lim_{\beta \to 0} \lim_{q\to 0}P^\text{1-loop}_{h,{\rm ind.}}(x^0,q)  
 \notag 
 \\
 &= \frac{2^{\frac{9}{4}} 3^{\frac{1}{2}} \left(\pi ^4-90 \,\zeta (5)\right)}{5^{\frac{3}{4}} \pi^{\frac{9}{2}}} (x^0)^3\left(\frac{H(x^0)}{M_{\rm pl}}\right)^3,\label{indresul}
\end{align}
where $\zeta(5)$ is the Riemann zeta function, $H(x^0)$ is the Hubble parameter at time $x^0$, and we have eliminated $\beta$ by using the Friedmann equation.
Further studies on the thermally induced GWs will be discussed elsewhere.

\subsection{$\mathcal O(\lambda^2)$: beyond mean field evolution}\label{sec:fis}

Next, we consider the $\mathcal O(\lambda^2)$ contribution in Eq.~\eqref{defOevol}.
We recast Eq.~\eqref{h23} into 
\begin{align}
		\delta \hat h^{(2,3)}_{ij}(x) &=\frac{4}{M_{\rm pl}^2}\int^{x^0}_{\tau_{\rm R}} dy^0  \int^{y^0}_{\tau_{\rm R}} dz^0 
		\int d^3 y
		\int d^3 z	\notag 
		\\
		&\times \partial^{(z)}_l \partial^{(y)}_m \frac{  \hat \chi(z) \hat \chi(y)
	+
	\hat \chi(y)  \hat \chi(z)}{2}
	\notag 
	\\
	&\times 
	\partial^{(y)}_n\partial^{(z)}_k  ia(z^0)^2 \left[  \hat \chi(y),   \hat \chi(z)\right]   
		\notag 
	\\
	&\times 
	\frac{M_{\rm pl}^2}{4}ia(y^0)^2 \left[\hat h_{ij}(x), \hat h^{mn}(y)\right]\hat h^{kl}(z) ,\label{lambda2ep3}
\end{align}
Let us compute the 1-loop order power spectrum from $\mathcal O(\lambda^2)$ field \eqref{lambda2ep3}. Since Eq.~\eqref{lambda2ep3} is $\mathcal O(\epsilon^3)$, we only have the CIS diagram at the 1-loop order. Similar to the CIS diagram at $\mathcal O(\lambda)$, the $\mathcal O(\lambda^2)$ order CIS diagram is calculated as 
\begin{align}
	{\rm Tr}\left[ \hat \varrho \hat h^{ij}\delta\hat h^{(2,3)}_{ij}\right] = {\rm Tr}\left[ \hat \varrho_h  \hat h^{ij} {\rm Tr}\left [\hat \varrho_\chi\delta\hat h^{(2,3)}_{ij}\right]\right].\label{traceorder}
\end{align}
Hence, we may evaluate the $\chi$ average first.
Using the scalar retarded Green functions~\eqref{rt:chi} and ~\eqref{rt:h}, and the Keldysh Green function~\eqref{kd:chi}, we find the $\mathcal O(\lambda^2)$ contribution
\begin{align}
	{\rm Tr}\left[\hat \varrho_\chi	\delta \hat h^{(2,3)}_{ij}(x) \right] &=\frac{4}{M_{\rm pl}^2}\int^{x^0}_{\tau_{\rm R}} dy^0 \int^{y^0}_{\tau_{\rm R}} dz^0 
		\int d^3 y
		\int d^3 z	
		\notag 
		\\
		&\times \partial_l^{(z)} \partial_m^{(y)} G^K(z,y) \partial_n^{(y)} \partial_k^{(z)}
		G(y,z) 
		\notag 
		\\
		&\times 
		G_{ij}{}^{mn}(x,y)\hat h^{kl}(z) \label{55}.
\end{align}
Eq.~\eqref{55} represents the radiation exchange of gravitational waves from $z$ to $x$.
The thermal state stimulates the radiation exchange, increasing gravitational wave amplitude at $x$.
Further simplification is possible in Fourier space.
We use the Fourier transforms of the Green functions~\eqref{green:scal}, \eqref{green:tens}, and \eqref{GKexp}, and then we integrate the delta functions.
Taking its thermal average, one finds
\begin{align}
	&{\rm Tr}\left[\hat \varrho_\chi	\delta \hat h^{(2,3)}_{ij}(x^0,\mathbf q) \right] 
	\notag 
	\\
	&= \frac{4}{M_{\rm pl}^2}\int^{x^0}_{\tau_{\rm R}} dy^0 \int^{y^0}_{\tau_{\rm R}} dz^0 
			\frac{\sin (q(x^0-y^0))}{qa(x^0)a(y^0)}\hat h^{kl}(z^0,\mathbf q)
		\notag \\
		&\times 
		\int \frac{d^3 p }{(2\pi)^3}
		 \int \frac{d^3p'}{(2\pi)^3}
		 		(2\pi)^3 \delta( \mathbf p + \mathbf p' -\mathbf q)
		 		\notag 
		 		\\
		 		&\times \Pi(\mathbf q)_{ij}{}^{mn} p'_{l}p'_{m}p_{n}p_{k}
		 		\notag 
		\\
		&\times
		\frac{\sin (p(y^0-z^0))}{p} \frac{\cos (p'(y^0-z^0))}{p' }f_{\beta p'}
			 ,\label{h23result}
\end{align}
where the projection tensor $\Pi(\mathbf q)_{ij}{}^{mn}$ is defined in appendix~\ref{propol}.
The angular dependence is more simplified after substituting this expression into Eq.~\eqref{traceorder}.
The 1-loop order power spectrum due to radiation exchange is defined as
\begin{align}
	&{\rm Tr}\left[\hat \varrho \left( \hat h^{ij}(x^0,\mathbf q')	\delta \hat h^{(2,3)}_{ij}(x^0,\mathbf q)
	+
	\delta \hat h^{(2,3)ij}(x^0, \mathbf q') \hat h_{ij}(x^0,\mathbf q)	 \right)\right] 
	\notag
	\\
	&=(2\pi)^3 \delta(\mathbf q+\mathbf q') P^\text{1-loop}_{h,{\rm ex.}}(x^0,q),\label{Pexchange}
\end{align}
can be evaluated as follows.
Substituting Eqs.~\eqref{h23result} into \eqref{Pexchange}, we find
\begin{align}
	P^\text{1-loop}_{h,{\rm ex.}}(x^0,q) &=\frac{16}{q^6} \left(\frac{H_{\rm inf.}^2}{M_{\rm pl}^2}\right)^2 \frac{\tau_{\rm R}^4}{(x^0)^2}  \int^{x^0}_{\tau_{\rm R}} \frac{dy^0}{y^0}  \int^{y^0}_{\tau_{\rm R}} \frac{dz^0}{z^0} 
	\notag 
	\\
	&\times 
	\sin (q(x^0-y^0))\sin(qx^0)\sin(qz^0) 
	\notag 
	\\
	&\times \mathcal J_\beta(q,y^0-z^0),\label{Pexchange1}
\end{align}
where $H_{\rm inf.}$ is the Hubble parameter during inflation, and we assumed instantaneous reheating, and we defined
\begin{align}
		\mathcal J_\beta(q,\tau) &\equiv  \int_0^{\infty} dp \int^{|p+q|}_{|p-q|} dp' \bar w(q,p,p')  
		\notag 
		\\
		&\times \frac{\sin (p\tau)}{p}\frac{\cos (p'\tau)}{p' }f_{\beta p'}.\label{Jdef22}
\end{align}

In the following, let us find an asymptotic formula for the IR gravitational waves.
In Eq.~\eqref{Jdef22}, the Planck distribution may be regarded as a window function that selects $p'\sim \beta$. 
The integrand is highly oscillating for $p'\tau \gg 1$.
Hence, the dominant contribution to $\mathcal J$ comes from ${p'}^{-1}\sim \tau \sim \beta$.

\begin{figure}[hbt]
  \includegraphics[width=\linewidth]{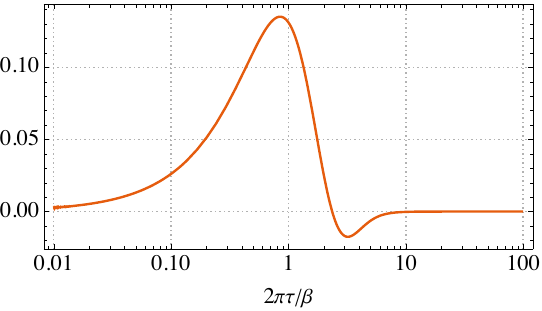}
  \caption{$\lim_{q\to 0} \beta^5\mathcal J_\beta(q,\tau)$ is plotted for $\tau$. The dominant contribution comes from $\tau\lessapprox \beta$.}
  \label{JIRfig}
\end{figure}

In the IR limit of the gravitational wave momentum, using Eq.~\eqref{IRform}, we find
\begin{align}
	&\lim_{q\to 0} \mathcal J_\beta(q,\tau) 
		\notag 
	\\
	&= 
	\frac{\pi ^3}{15 \beta^5} \left(11 \cosh \left(\frac{2 \pi  \tau }{\beta}\right)+\cosh \left(\frac{6 \pi  \tau }{\beta}\right)\right) \text{csch}^5\left(\frac{2 \pi  \tau }{\beta}\right)
	\notag 
	\\
	&-\frac{1}{40 \pi ^2 \tau ^5}.\label{eq:Jwindow}
\end{align}
Note that Eq.~\eqref{eq:Jwindow} is valid for $q\tau\sim  q\beta \sim q/p' \ll1 $.
Namely, the approximation is valid as long as the gravitational wave wavelengths are sufficiently longer than those of photons, not restricted to the super Hubble scales.
It is convenient to change the integral order as
\begin{align}
	\int^{x^0}_{\tau_{\rm R}} dy^0  \int^{y^0}_{\tau_{\rm R}} dz^0 = \int^{x^0}_{\tau_{\rm R}} dz^0 \int^{x^0}_{z^0} dy^0.
\end{align}
As presented in Fig.~\ref{JIRfig}, since $\mathcal J_\beta(q,y^0-z^0)$ is non-vanishing only in the range,
\begin{align}
	y^0-z^0 \lessapprox \beta \sim \sqrt{\frac{H}{M_{\rm pl}}}\frac{1}{aH} \ll \frac{1}{aH},
\end{align}
the contribution to the integral comes only from $z^0\sim y^0$.
Thus, the radiation exchange in the very short time interval contributes to the correction,
hence we may approximate the $y^0$ integral as
\begin{align}
	&\int^{x^0}_{z^0} \frac{dy^0}{y^0} \sin (q(x^0-y^0)) \mathcal J_\beta(q,y^0-z^0)
	\notag 
	\\
	&\sim 
	\frac{\sin (q(x^0-z^0))}{z^0} \int^{x^0}_{z^0} dy^0  \mathcal J_\beta(q,y^0-z^0).\label{y0int}
\end{align}

In the IR limit of the gravitational waves and the high-temperature limit~($\beta\ll x^0-z^0\sim x^0$, as the radiation exchange happens at an earlier time.), we find an asymptotic expression for the $y^0$ integral:
\begin{align}
	&\int^{x^0}_{z^0} dy^0  \mathcal J_\beta(q,y^0-z^0) 
	\notag \\
	=&\frac{1}{7200 \pi ^2 \beta ^4}\Bigg[16 \pi ^4 + \frac{45 \beta ^4}{(x^0-z^0)^4}
	\notag 
	\\
	&-240 \pi ^4 \left(\cosh \left(\frac{4 \pi  (x^0-z^0)}{\beta }\right)+2\right) \notag 
	\\
	&\times\text{csch}^4\left(\frac{2 \pi  (x^0-z^0)}{\beta }\right)\Bigg].
\end{align}
Note that $\beta$ is the inverse of the comoving temperature comparable to the comoving wavelength of typical thermal photons. 
The high-temperature limit means the hierarchy expressed in Eq.~\eqref{hightemplim}, i.e., the typical wavelength of photons is negligibly smaller than the size of the universe.
Thus, using the fact that $\beta$ is the fastest time scale, we obtain
\begin{align}
	\lim_{\beta\to 0}\lim_{q\to 0}\int^{x^0}_{z^0} dy^0  \mathcal J_\beta(q,y^0-z^0) =  \frac{\pi^2}{450\beta^4}\label{Jintnew}
\end{align}
Combining this result with Eqs.~\eqref{Pexchange1} and \eqref{y0int}, we find the following asymptotic formula: 
\begin{align}
	& \lim_{\beta \to 0} \lim_{q\to 0} P^\text{1-loop}_{h,{\rm ex.}}(x^0,q) \notag 
	\\
	&=\frac{16}{q^3} \left(\frac{H_{\rm inf.}^2}{M_{\rm pl}^2}\right)^2  \frac{\tau_{\rm R}^4 \pi^2}{450\beta^4}    \left(\frac{\tau_{\rm R}}{x^0} + \log\left(\frac{x^0}{\tau_{\rm R}}\right)-1 \right).
\end{align}
As the thermal scalar field dominates the Universe as a radiation fluid, by using the Friedmann equation, one finds
\begin{align}
	\frac{\tau_{\rm R}^4 \pi^2}{450\beta^4} = \frac{1}{15}  \frac{\pi^2}{30a(\tau_{\rm R})^4 \beta^4} a(\tau_{\rm R})^4 \tau_{\rm R}^4  = \frac{1}{5}\frac{M_{\rm pl}^2}{H_{\rm inf.}^2}.\label{taurhubble}
\end{align}
To summarize, we find
\begin{align}
	P^\text{1-loop,IR}_{h,{\rm ex.}}(x^0,q)  =\mathcal F\left(\frac{x^0}{\tau_{\rm R}}\right) P^{\rm IR}_{h0}(x^0,q),\label{result2}
\end{align}
where we defined
\begin{align}
	P^{\rm IR}_{h0}(x^0,q) & = \frac{4}{q^3} \frac{H_{\rm inf.}^2}{M_{\rm pl}^2},\label{PGWPower}
	\\
	\mathcal F(X) &= \frac{4}{5}\left( \log X-1 + \frac{1}{X} \right).
\end{align}
Note that $\mathcal F(X)$ is a monotonic function of $X$ for $X\geq 1$, and $\mathcal F(1)=0$. 
$F(X)$ exceeds unity for a few e-folds.
In the actual Universe, the maximum value of $x^0$ is the conformal time at matter radiation equality. We do not know the reheating time $\tau_{\rm R}$, but the ratio $x^0/\tau_{\rm R}$ can be typically 50 to 60 e-folds in our Universe. In this case, the enhancement factor is not small:
\begin{align}
	\mathcal F(x^0/\tau_{\rm R}) = \mathcal O(10),\label{eq83}
\end{align}
which implies that the 1-loop correction exceeds unity in the radiation-dominant Universe.
Unlike the local 1-vertex contribution, the nonlocal 2-vertex diagram is not canceled by the tadpole.
However, we need to justify the result of the perturbative analysis more carefully.

To summarize, the one-loop radiation exchange diagram results in a scale-independent enhancement of the primordial gravitational waves on super-horizon scales. In the calculation above, we applied various limits to obtain analytic results. Below, we summarize the hierarchies assumed in the analysis.
First, all dimensionful parameters, except for the Hubble parameter, are comoving ones, and the physical values are to be obtained by multiplying them by the scale factor. 
The fastest timescale is the inverse temperature $\beta$ in the Planck distribution function, which also constrains the typical wavelength of photons as $\beta \sim p'^{-1}$.
We introduced three different conformal times: $x^0$, $y^0$, and $z^0$. Here, $x^0$ is the conformal time when the gravitational waves are evaluated, while $y^0$ and $z^0$ are integration variables that represent the conformal times for the photons in the exchange diagram.
The exchange process is expressed using properly time-ordered Green functions, leading to the relation $x^0 \geq y^0 \geq z^0 \geq \tau_{\rm R}$, where $\tau_{\rm R}$ is the reheating time related to the inflationary Hubble parameter $H_{\rm inf}$ by Eq.~\eqref{taurhubble}, assuming instantaneous reheating.
Then we find that the $y$-integral in Eq.~\eqref{Jintnew} is dominated by $y^0 - z^0 \sim \beta$, implying that the radiation exchange effect is dominated by a very short time interval. 
Finally, we integrate over $z^0$ from the initial time $\tau_{\rm R}$ to $x^0$ to obtain the final result.

To derive the analytic IR formula, we imposed the condition $q x^0 \ll 1$, corresponding to the super-horizon scales. However, a similar result is obtained by taking $x^0 \gg z^0$ and $q \beta \ll 1$, as assumed in Eq.~\eqref{eq:Jwindow}. In this case, the sine functions remain, indicating that the enhancement is also present for the propagating mode on sub-horizon scales, while $x^0\sim z^0$ contribution must be more carefully considered. Therefore, the hierarchy between the GW wavelength and the photon wavelength is crucial in this calculation.

%%%%%%%%%%%%%%%%%%%%%% MS Oct 17

It is worth noticing that in addition to our calculation here, the structure of $\mathcal F(X)$ can also arise in another situation.
We may consider a source proportional to a constant tensor fluctuation $h^0_{ij}$ in the RHS of the classical equation of motion~\eqref{eomtens}:
\begin{align}
    \xi_{ij} = \frac{4}{5}a^2H^2 h^0_{ij}.\label{randomforce}
\end{align}
The IR solution in the presence of Eq.~\eqref{randomforce} is written in the form of 
\begin{align}
    h_{ij} = \mathcal F(x^0/\tau_{\rm R}) h^0_{ij},
\end{align}
after fixing the two integral constants to satisfy $h_{ij}(\tau_{\rm R})=h'_{ij}(\tau_{\rm R})=0$.
However, this solution \textit{cannot} be regarded as a gauge mode due to a constant gauge transformation, 
\begin{align}
    h_{ij} \to h_{ij}+ h^0_{ij}.\label{constgauge}  
\end{align}
This is induced by an infinitesimal coordinate transformation,
\begin{align}
   \xi^i = \frac{1}{2}\delta^{ik}x^j h^{0}_{kj}, 
\end{align}
with an IR constant mode $h^0_{ij}$.
The gauge transformation~\eqref{constgauge} of Eq.~\eqref{chiichij} yields a shift $a^2P_\gamma h^0_{ij}$ in the RHS, and thus a source term like Eq.~\eqref{randomforce} in Eq.~\eqref{eomtens:2}. However, the mass term in the LHS is also shifted by the same amount by the gauge transformation, so the net contribution is zero. Thus, the linear equation of motion is manifestly gauge invariant so that the source term cannot arise due to the gauge transformation.

So far, we have considered that the transition from inflation to radiation dominant is instantaneous, and de Sitter approximation is assumed for inflation.
If the Hubble parameter varies from the inflationary Hubble $H_{\rm inf.}$ to $H_{\rm R}$ during non-instantaneous reheating, the Hubble-to-Planck mass ratio in Eq.~\eqref{Pexchange1} is replaced as
\begin{align}
	\left(\frac{H_{\rm inf.}^2}{M_{\rm pl}^2}\right)^2 \to \left(\frac{H_{\rm R}^2}{M_{\rm pl}^2}\right)\left(\frac{H_{\rm inf.}^2}{M_{\rm pl}^2}\right).
\end{align}
However, Eq.~\eqref{result2} is unchanged since the $H_{\rm R}$ dependence vanishes in the end.

\section{Higher loop corrections}
\label{sec:higher}

In the previous section, we showed that the 1-loop correction may exceed the tree-level spectrum, implying that the spectrum's loop expansion might not be convergent. 
In this section, we examine the issue more carefully by focusing on the expansion in the Hubble-to-Planck mass ratio, $\alpha \equiv H_{\rm inf.}/M_{\rm pl} \ll 1$. 
The 1-loop correction is $\mathcal{O}(\alpha^2)$, comparable to the linear power spectrum. Higher-order loop spectra also contain $\mathcal{O}(\alpha^2)$ terms. Thus, the order of loops does not align with the order in $\alpha^2$, unlike, for example, the inflationary loops~\cite{Senatore:2009cf}. Hence, reorganizing the loop expansion in terms of $\alpha$ may be important. This section discusses how to count the order in $\alpha$ from a diagrammatic perspective.

We will show that 2-loop or higher-order corrections are needed to fully capture $\mathcal{O}(\alpha^2)$ contributions, which are beyond the scope of this paper. We provide only a sketch of the calculation and defer the actual computations to future work.
We can estimate the order in $\alpha$ as follows. Schematically, the quantum correction to the tensor perturbation at the $(n,m)$ order, where $n$ is the order in the interaction Hamiltonian and $m$ is the order in the amplitude of the perturbation, is expressed as
\begin{align}
    \delta \hat{h}^{(n,m)} = \int \left[\partial^2 G \right]^{n-k} \left[\frac{G_T}{M_{\rm pl}^2} \right]^{\frac{m-\ell}{2}} \hat{h}^\ell (\partial \hat{\chi})^{2k} ,\label{schematic}
\end{align}
where $G$ and $G_T$ are the scalar and tensor Green functions, respectively. Note that $G$ and $G_T$ appear after simplifying the commutators in Eq.~\eqref{defOevol}. Here, $m-\ell$ is always an even integer, with $\ell$ being the number of uncontracted tensor fluctuations $\hat{h}$. Similarly, $k$ counts the number of uncontracted scalar fields $\hat{\chi}$.

Since the enhancement factors must be dimensionless, there should be a dimensionful factor that compensates for the dimensions given by the powers of $M_{\rm pl}$. We expect this factor to be in powers of $H_{\rm inf}$, particularly in the IR limit, resulting in ${\cal O}(M_{\rm pl}^{-n}) = {\cal O}(\alpha^n)$. While an explicit confirmation of this expectation is eventually necessary, it is beyond the scope of this paper.

Let us first briefly count the powers of $M_{\rm pl}$ in Eq.~\eqref{schematic}.
The second derivative of the scalar Keldysh Green function, $\partial^2 G^K$, is $\mathcal{O}(\beta^{-4}) = \mathcal{O}(M_{\rm pl}^2)$ at finite temperature. 
The uncontracted scalar field is typically evaluated as $\partial \chi \sim \sqrt{\partial^2 G^K}$.
Hence, positive powers of $M_{\rm pl}$ appear in the power spectrum. This contrasts with the inflationary case, where the statistical average is taken over the vacuum state and no positive powers of $M_{\rm pl}$ appear. As a result, the $n$-loop diagram is $\mathcal{O}(M_{\rm pl}^{-2n-2})$ in the inflationary case. In contrast, in the finite temperature case, $n$-loop diagrams acquire additional powers of $M_{\rm pl}$.
Typically, the uncontracted scalar will introduce $M_{\rm pl}^{2k}$ in Eq.~\eqref{schematic}.

Thus, the finite temperature effects shift the expansion hierarchy in terms of $M_{\rm pl}$ due to the many excited fields involved in the interaction. 
Higher-order loop calculations are more complicated than the 1-loop case, so they are beyond the scope of this paper. 
Here, we provide a sketch of the two-loop contributions.

The order counting of the Green functions and uncontracted fields in terms of the Planck mass is given as:
\begin{align}
    \partial^2 G &= \includegraphics[width=2cm]{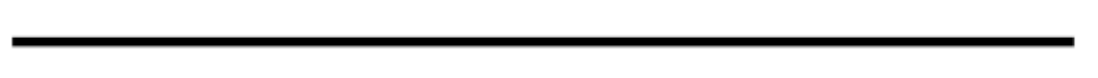} = \mathcal{O}(M_{\rm pl}^0), \label{pa2G} \\
    \frac{G_{T}}{M_{\rm pl}^2} &= \includegraphics[width=2cm]{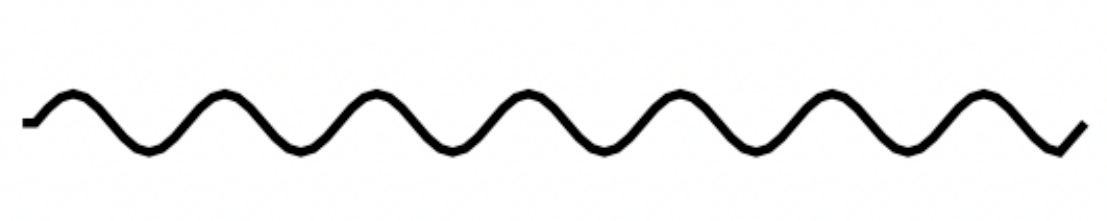} = \mathcal{O}(M_{\rm pl}^{-2}), \label{me2G} \\
    \partial \hat{\chi} &= \includegraphics[width=2cm]{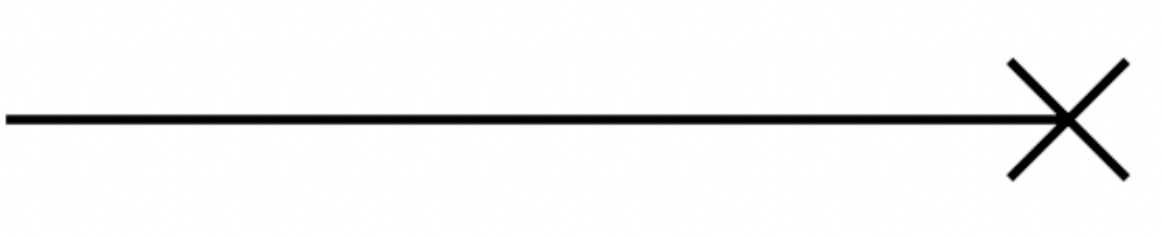} = \mathcal{O}(M_{\rm pl}), \label{chi_al} \\
    \hat{h}_{ij} &= \includegraphics[width=2cm]{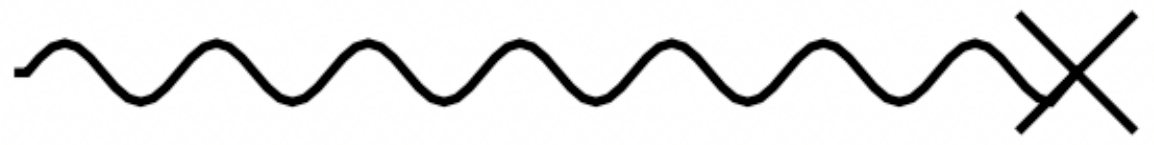} = \mathcal{O}(M_{\rm pl}^{-1}), \\
    \int d^4x &= \bullet = \mathcal{O}(M_{\rm pl}^0). \label{H4dV}
\end{align}
Eq.~\eqref{chi_al} implies a thermal average, i.e., $\partial \hat{\chi} = \mathcal{O}(\sqrt{\partial^2 G^K})$. 
The $M_{\rm pl}^{-2}$ factor in Eq.~\eqref{me2G} comes from normalizing the tensor retarded Green function, ensuring that it always appears alongside the Green function (see, e.g., Eq.~\eqref{eqh12}).
All relevant terms in the 1-loop diagrams are given as:
\begin{widetext}
\begin{align}
    \delta \hat{h}^{(1,2)}_{ij}(x) &= i \int^{x^0} dy^0 [\hat{H}_{h\chi\chi}(y^0), \hat{h}_{ij}(x)] = 
    \begin{gathered}
    \includegraphics[width=2.5cm]{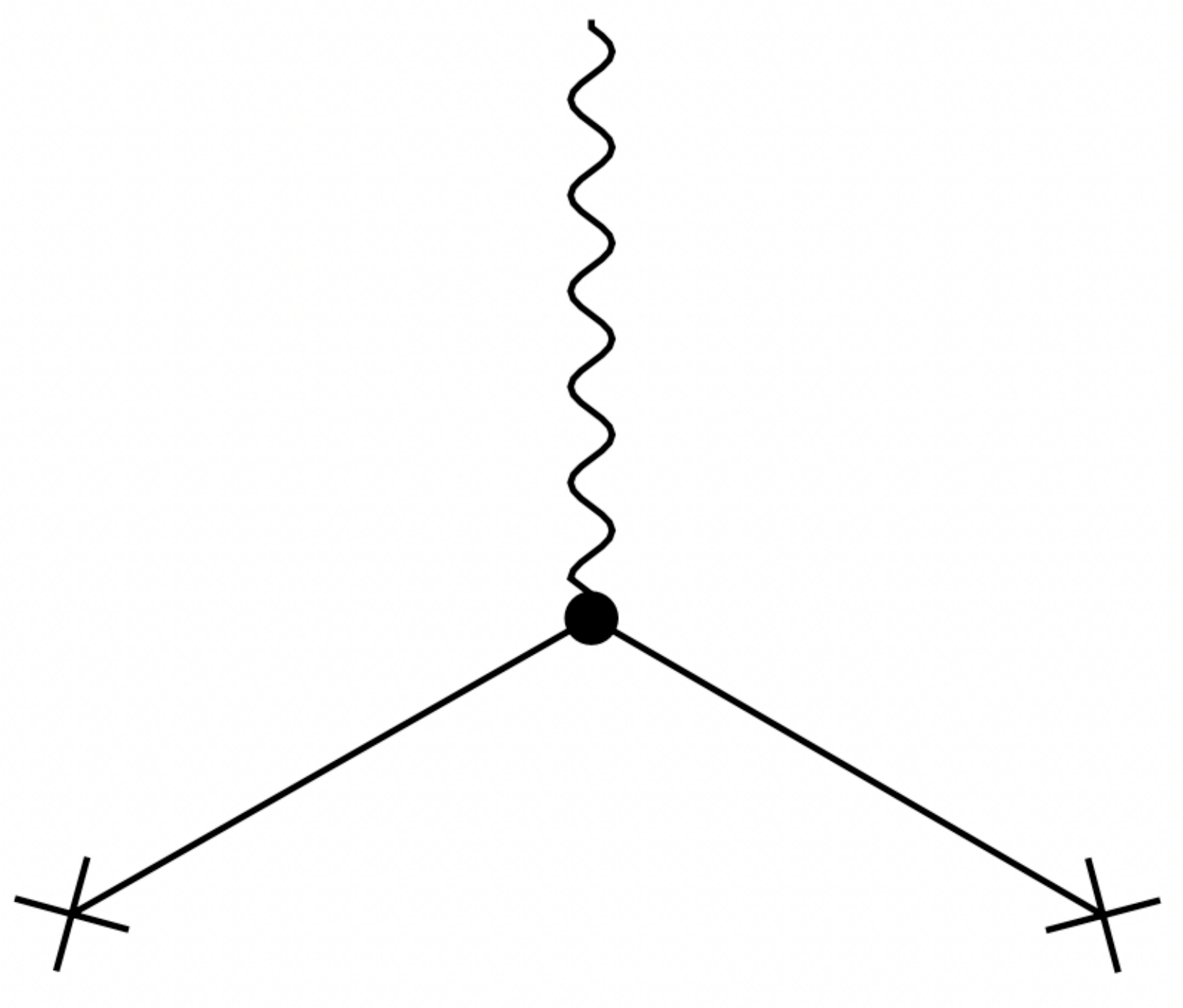}	
    \end{gathered}
	 = \mathcal{O}(M_{\rm pl}^0), \label{digh12}\\
    \delta \hat{h}^{(1,3)}_{ij}(x) &= i \int^{x^0} dy^0 [\hat{H}_{hh\chi\chi}(y^0), \hat{h}_{ij}(x)] =
    \begin{gathered}
	     \includegraphics[width=2cm]{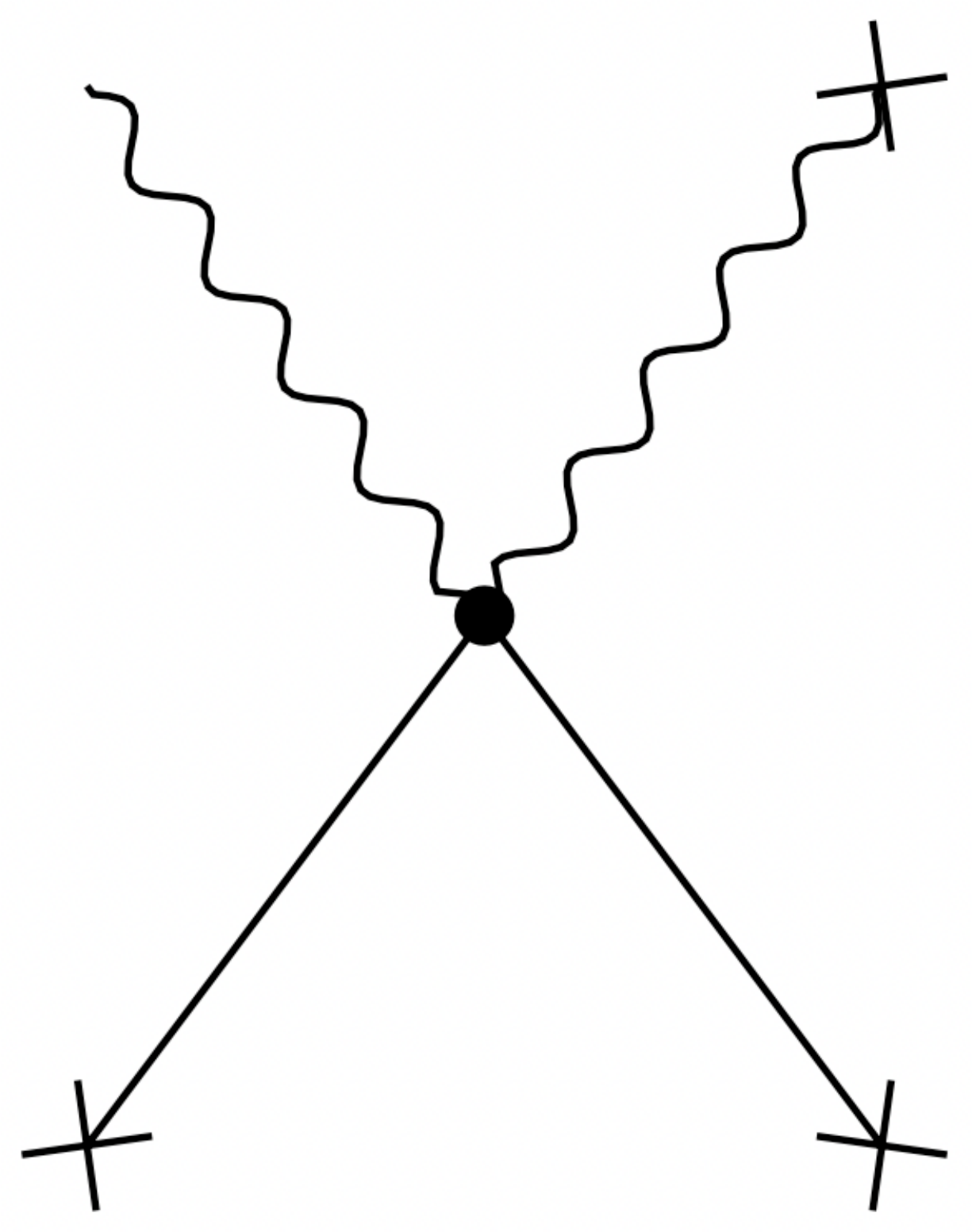}    	
    \end{gathered}
 = \mathcal{O}(M_{\rm pl}^{-1}), \\
    \delta \hat{h}^{(2,3)}_{ij}(x) &= i \int^{x^0} dy^0 i \int^{y^0} dz^0 \left[\hat{H}_{h\chi\chi}(z^0), \left[\hat{H}_{h\chi\chi}(y^0), \hat{h}_{ij}(x)\right]\right] = 
    \begin{gathered}
    	\includegraphics[width=3cm]{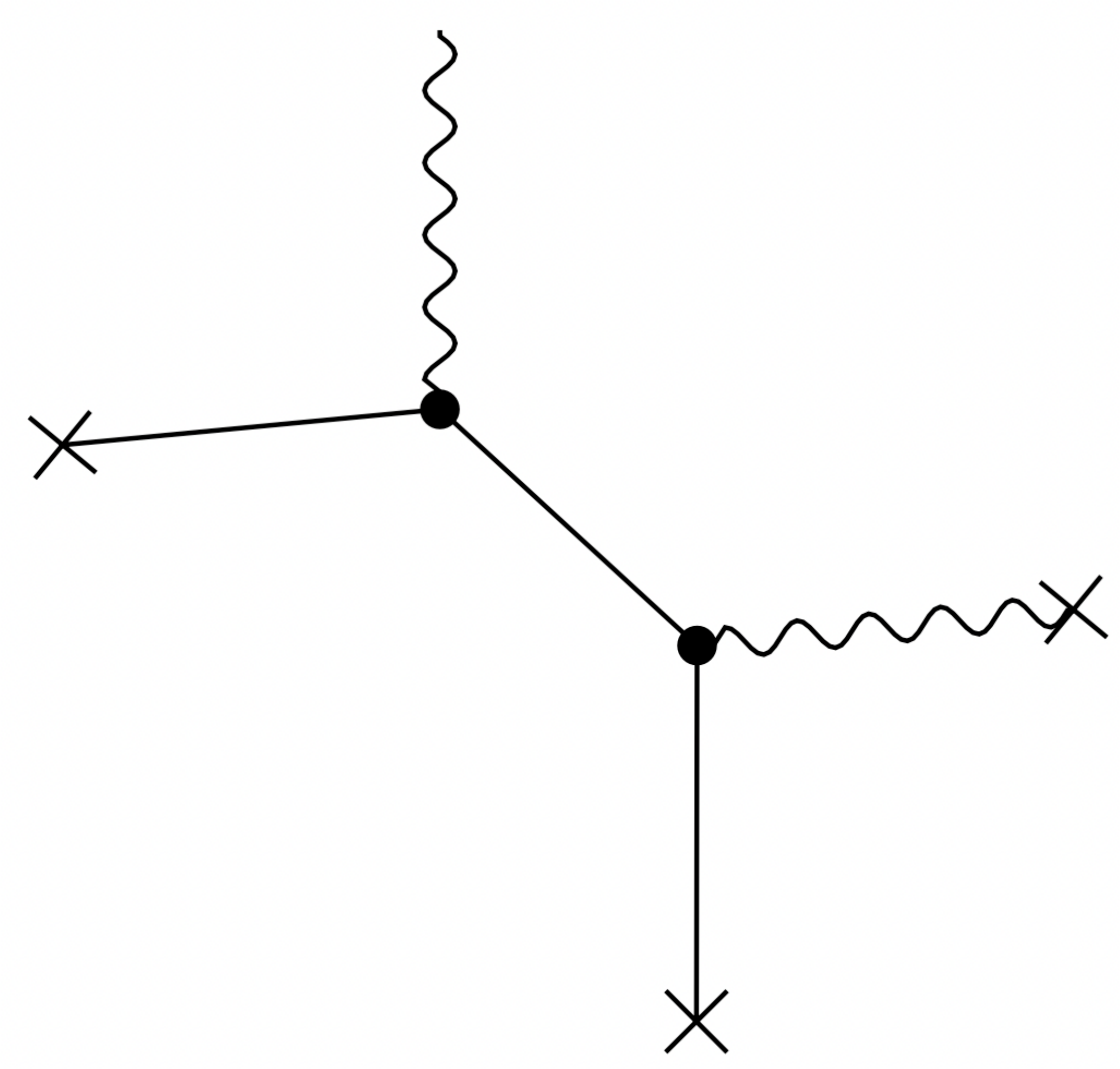}
    \end{gathered} = \mathcal{O}(M_{\rm pl}^{-1}).\label{diagh23}
\end{align}
\end{widetext}

Higher-order loops are generated by these diagrams iteratively as one can write
\begin{align}
	\delta \hat h^{(n+1,m+\ell)} (x)=\int^{x^0}  dy^0 [\hat H_{\underbrace{h\cdots h}_{\ell~{\rm tensors}}\chi\chi}(y^0),\delta \hat h^{(n,m)}(x) ],
\end{align}
where $\hat H_{h\cdots h\chi\chi}$ is the interaction Hamiltonian with $\ell$ graviton vertices, defined similar to Eqs.~\eqref{Hhchichi} and \eqref{Hhhchichi}, and the order of integration in time is rearranged.
As we have the commutator, one of the uncontracted fields in $\delta \hat h^{(n,m)}$ is contracted with a field in the interaction Hamiltonian.
Diagrammatically, we replace an external source with a sub-diagram obtained from the interaction Hamiltonian.
For instance, $\delta h^{(2,3)}_{ij}$ generates $\delta h^{(3,4)}_{ij}$ by adding the sub-diagram ~\eqref{digh12} to the diagram~\eqref{diagh23}:
\begin{widetext}
\begin{align}
    \delta \hat{h}^{(3,4)}_{ij}(x) = 
    \begin{gathered}
    	\includegraphics[width=10cm]{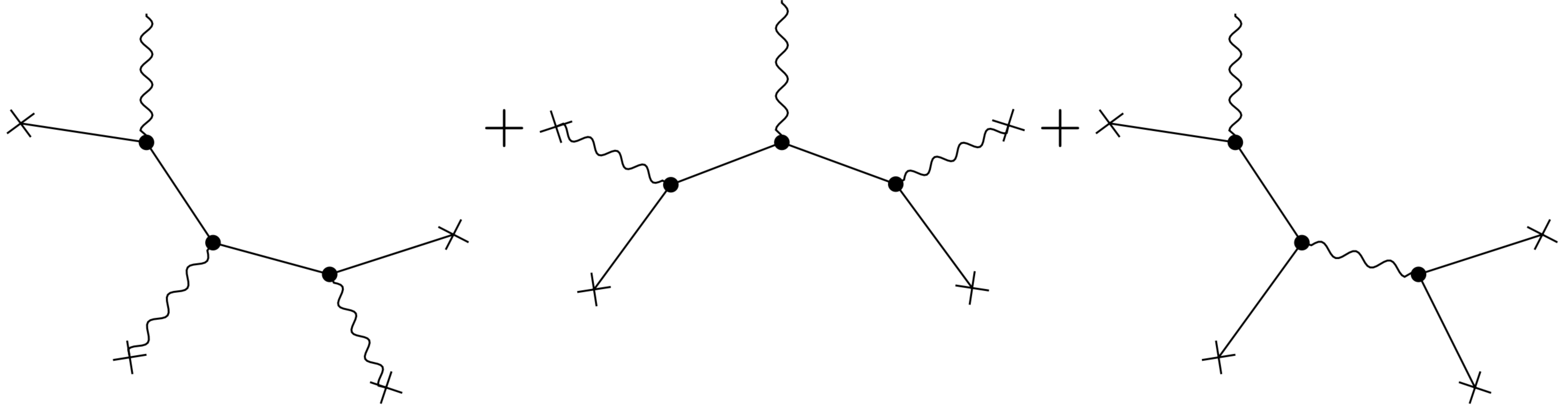}.
    \end{gathered}
\end{align}
\end{widetext}
In this way, we can draw arbitrary higher-order corrections systematically.
For the 2-loop CIM graphs, we need to consider contributions up to $\delta \hat{h}^{(3,4)}_{ij}$. 
The 2-loop CIS diagrams are more complicated, requiring the inclusion of terms like $\delta \hat{h}^{(4,5)}_{ij}$.

Finally, we should mention that thermal state corrections, which effectively add tadpoles to the spectrum, have not been included in these multi-loop diagrams. 
The cancellation of tadpole diagrams, as seen in the 1-loop case (e.g., diagrams (a) and (b) in Fig.~\ref{fnpow}), may occur. 
As thermal state corrections introduce extra tensor legs to the scalar field density operator, any corrected diagrams will have an additional factor of $\mathcal{O}(M_{\rm pl}^{-1})$. 
Therefore, the tadpole diagram capable of canceling an $\mathcal{O}(M_{\rm pl}^{-2})$ power spectrum is, e.g., the diagram (a) in Fig.~\ref{fnpow}, which is $\mathcal{O}(M_{\rm pl}^{0})$.

\section{Conclusions}
\label{sec:conc}

In the literature, radiation was regarded as a smooth and continuous fluid in a scenario of the radiation-dominated universe.
The quantum nature of the field was averaged over the universe so that the radiation only drives the cosmic expansion from a macroscopic perspective. 
This paper considers interactions between gravitational waves and individual massless elementary particles of the fluid.
Thermal radiation is composed of a large number of excited state fields.
Therefore, radiation exchange may be important for gravitational waves from a quantum field theory point of view.
We evaluated the gravitational wave power spectrum at the 1-loop level using the in-in formalism, based on the full action of the radiation-dominant universe before course-graining the constituents.

We found two kinds of 1-loop contributions in the spectrum: the local 1-vertex diagram (Fig~\ref{fnpow}(b)) and the nonlocal 2-vertex diagrams (Figs~\ref{fnpow}(c) and (d)).
The 1-vertex diagram comprises the local vertex connected to the radiation and the gravitational waves, which the equivalence principle should eliminate.
We found that the perturbed tadpole diagram (Fig~\ref{fnpow}(a)) exactly cancels this contribution.
The cancellation of the 1-loop local vertex diagram is consistent with the massless linearized Einstein equation for classical gravitational waves, which implies that the classical Einstein equation can be obtained by taking the mean-field approximation of the Heisenberg equation.
Nonlocal vertex diagrams are further classified into two types: the thermally induced gravitational waves (Fig~\ref{fnpow}(c)) and the radiation exchange (Fig~\ref{fnpow}(d)).
The local coordinate transformation does not generally eliminate these since they contain gravitational waves at different spacetime locations.
We evaluated these diagrams in the high-temperature limit and found some analytic formulas. 
The thermally induced gravitational waves may be created by the random inhomogeneous configuration of the scalar field, which can be suppressed in the IR scale.

On the other hand, the radiation exchange diagram leads to an enhancement of IR gravitational waves. 
It is important to note that the term "IR scale" does not necessarily refer to the super-horizon scale. 
Our computations are valid as long as the wavelengths of gravitational waves are significantly longer than those of photons.
 
The 1-loop radiation exchange amplifies the primordial gravitational waves by a factor of $10$, implying that the reorganization for the loop expansion may be needed. 
Note that the tadpole cancellation does not happen to this contribution, at least at the 1-loop level.

We pointed out that the loop expansion can be reorganized by the power of $\alpha \equiv H_{\rm inf.}/M_{\rm pl}$, generally a small parameter in the radiation-dominant universe.
Actual calculations are postponed to future works, but we presented a sketch of higher order loop calculation in a diagrammatic perspective. 
For the consistent calculation to $\mathcal O(\alpha^2)$, we need to evaluate all of the diagrams. 
 Also, it remains to be seen whether the scalar perturbation of the metric mixes in at the two-loop order.
Similar IR classical 1-loop corrections in the context of primordial black hole formation were discussed in Refs.~\cite{Ota:2020vfn,Chen:2022dah}.

At first sight, a concern may be raised about the causality in the enhancement of super-horizon tensor fluctuations. 
The diagram (c) in Fig.~\ref{fnpow} describes the gravitational waves induced by causal interactions in thermal plasma. As they are induced gravitational waves, they are suppressed on super-horizon scales due to causality, as shown in Eq.~\eqref{indresul}.
On the other hand, the radiation exchange is a correction to the primordial gravitational waves that already existed from the beginning of the radiation-dominated era, presumably produced during inflation. Namely, the radiation exchange does not generate a new super-horizon correlation: It is a correction to the propagation of gravitational waves.
Also note that the exchange diagrams are constructed from properly time ordered Green functions, and hence they are free from any causality violations.
One of the ways to understand the effect is to note that the super horizon evolution can be considered as a Bogoliubov transformation~\cite{Ota:2022hvh, Ota:2022xni}, which is the time evolution of the vacuum state brought by causal interactions. 

We also point out the similarity between the radiation exchange and the stimulated emission, the mechanism of laser. In fact, the relation between the spectra of the induced gravitational waves obtained in Sec.~\ref{sec:mf} and of radiation exchange in the IR limit resembles that of the induced emission and stimulated emission.
Further investigation into this similarity may shed more light on the effects of radiation exchange.

Gravitational waves from the standard model particle plasma in the early universe were also considered in Refs.~\cite{Ghiglieri:2015nfa,Ghiglieri:2020mhm,Ringwald:2020ist}, with a specific focus on thermally induced gravitational waves. In contrast, the radiation exchange was not considered.
They considered the mean-field approximation for gravitational waves, while loops of standard model fields are considered in the energy-momentum tensor.
 
The loop expansion counts the order in $\alpha^2$ for the inflationary loops because the scalar Keldysh Green function is also $\mathcal O(\alpha^2)$ for the vacuum state. 
The shift in the expansion hierarchy happens in our case because many excited particles stimulate the radiation exchange.

We have not introduced any nonstandard physics in our calculation, while we regarded the radiation field as a massless scalar field for simplicity.
Extension to the standard model particles is straightforward, but the result will not change drastically since only the thermal nature is crucial in our calculation.
The radiation exchange should exist unless they are exactly canceled up to the 2-loop order for some reason.
Hence, further qualitative analysis is crucial. 

We mention several possibilities for canceling the IR correction we obtained.
We showed that the local tadpole eliminates the single vertex 1-loop diagram, resulting from the equivalence principle.
There might be missing tadpoles to nonlinear order, and the equivalence principle may also eliminate the radiation exchange diagram.
In particular, the high-temperature limit implies that the two vertices in the diagram (d) in Fig~\ref{fnpow} are close.
In this limit, the two vertices reduce to a single local vertex, which the equivalence principle might eliminate, while we do not expect such a counter-term. % in the diagrams in Fig.~\ref{fig2}. 
Cancellation, including the 2-loop order, may also happen.
Recent studies on the 1-loop effect for the ultra-slow roll~(USR) single field inflation~\cite{Kristiano:2022maq} reported a similar nonlinear correction.
They found that the loop effect during the USR phase might largely amplify the IR curvature perturbation during inflation. Thus, the instability of perturbative expansion might be an issue.
Similar effects on gravitational waves are also discussed in Ref.~\cite{Firouzjahi:2023btw}. 
The claim was controversial as the conservation of IR curvature perturbation and stability of perturbations are crucial in cosmological perturbation theory.
They discussed inflationary loops from the vacuum fluctuation, which differs from our setup as we focus on the finite temperature effect in the radiation-dominant universe.
However, the IR feature looks quite similar to our case.
There are several objections to the claim of Ref.~\cite{Kristiano:2022maq}, which may also apply to our case.
Ref.~\cite{Riotto:2023gpm} argued that the smooth transition from slow roll to USR phase may eliminate the loop corrections~(see also Refs.~\cite{Kristiano:2023scm,Ballesteros:2024zdp,Braglia:2024zsl,Choudhury:2024jlz,Inomata:2024lud,Firouzjahi:2024psd,Iacconi:2023ggt,Davies:2023hhn,Maity:2023qzw,Fumagalli:2023hpa,Cheng:2023ikq,Franciolini:2023lgy,Firouzjahi:2023ahg} for recent progress.).
In our case, we considered the de Sitter approximation for the inflationary phase and assumed that the transition from inflation to radiation dominant is instantaneous.
Hence, the inflationary Hubble parameter is identified with that at reheating, $H_{\rm R}$ in the present case.
Depending on the models for reheating and inflation, we may find some different coefficients as the Hubble parameter is not generally constant.
Ref.~\cite{Fumagalli:2023hpa} considered the boundary term in the interaction Hamiltonian exactly cancels the loop correction in Ref.~\cite{Kristiano:2022maq}.
In our case, we only have the kinetic term of the scalar field, and thus, no boundary term appears in the interaction. Recently, Ref.~\cite{Tada:2023rgp} also claimed that the one-loop effect on the IR curvature power spectrum is totally eliminated from Maldacena's consistency relation, but they also confirmed that the same argument does not apply to the tensor power spectrum.
One may wonder if the diagram (b) in Fig.~\eqref{fnpow} is canceled by (d) instead of the tadpole (a), as (d) can be seen as the negative mass squared in Eq.~\eqref{randomforce}. However, this kind of cancellation will never happen as (b) is a field redefinition dependent while (d) is not.

We also would like to make a comment from a hydrodynamical perspective.
We saw that the IR correction could be considered as an IR solution in the presence of a constant stochastic noise~\eqref{randomforce} on the right-hand side of the classical equation of motion~\eqref{eomtens}. 
This effect could be interpreted as a next-to-leading order correction in the covariant gradient expansion because it is of $\mathcal O(H^2)$.
In this regard, it may be noted that the number of the hydrodynamical coefficients is often reduced by the field redefinition of fluid velocity and temperature, i.e., by a change of frames. 
However, the consistency between the background dynamics and the radiation equation of state must be carefully studied before we make a conclusive statement.

Another interesting aspect is comparing our result to the textbook treatment of IR divergence in flat space QFT. In flat space QFT, the soft Bremsstrahlung of IR gravitons suffer an IR divergence, which can be resummed, and IR-safe observables should be defined. However, in our case, the expanding universe and the Keldysh propagator make the behavior of IR gravitons tricky. Since the IR divergence of soft tensor modes does not occur, and the long wavelength tensor modes should be, in principle, observable when they return to the horizon, no special IR treatment is done in the present work. Also, our calculation has no IR divergence in the scalar loop, thanks to the expanding background. Having that said, it remains interesting to understand whether similar IR resummations can be done even if the calculation is not divergent. 

There remains, however, a fundamental issue related to the very meaning of the tensor field on superhorizon scales. As pointed out in Ref.~\cite{Domenech:2020xin}, a static transverse traceless component of the spatial metric can be gauged away at second-order gauge transformation of the time slicing.
Thus, one could, in principle, perform such a transformation to remove the tensor field at the IR limit. 
Then the remaining tensor field would vanish in the IR limit. 
If one assumes this remaining part to be the one the scalar field interact with, the 1-loop correction would be correspondingly suppressed.

Relating to this issue, it was also pointed out in Ref.~\cite{Domenech:2020xin} that computing the induced gravitational waves in the synchronous gauge can be tricky. 
Namely, if one computes the induced gravitational waves due to a scalar perturbation in the synchronous gauge, the resulting amplitude turns out to contain a very large gauge mode when it re-enters the Hubble horizon. 
This gauge mode must be removed before one evaluates the amplitude of the induced gravitational waves on small scales inside the Hubble horizon.
This is due to the fact that the synchronous gauge does not uniquely fix the time slicing. 
In this paper, since we assumed there is no scalar perturbation on large scales, perhaps this issue is irrelevant. Nevertheless, we should regard this as a caution when dealing with the tensor field in the IR limit.

Regarding the renormalization issue, the divergence may stem from the integrals in the loop analysis due to the vacuum fluctuations. In the present calculation, such divergence is absent since we are focusing on the contribution of thermally excited modes, which are exponentially suppressed in the UV limit.
It may be of interest to consider vacuum loop contributions as discussed in Ref.~\cite{Landsman:1986uw}.  However, it is a separate issue.
While total consistency may be ideal, it obscures the essence if we demand unnecessary details when discussing another layer of the problem.
Therefore, throughout the calculation, we dropped the zero temperature contribution, assuming the regularization and renormalization prescriptions for the zero temperature contribution have been properly applied. 

Finally, we mention that gravitational wave self-interactions are ignored for simplicity. It is justified because the Keldysh Green function for gravitational waves is not excited like the radiation, so the 1-loop correction will be at most $\mathcal O(\alpha^4)$ after proper renormalization procedures.

\begin{acknowledgments}
We would like to thank Junsei Tokuda for his critical comments about the two-loop order analysis in the previous version of our manuscript.
We would like to thank Chen Chao, Huiyu Zhu, and Yuhang Zhu for carefully reading our manuscript and valuable comments.
We thank Andrew Cohen, Tsutomu Kobayashi,  Toshifumi Noumi, Zhong-Zhi Xianyu, Masahide Yamaguchi, and Richard Woodard for useful discussions. This work is supported in part by the Natural Science Foundation of China under Grant No.12147102. AO and YW are supported in part by the National Key R\&D Program of China (2021YFC2203100), the NSFC Excellent Young Scientist Scheme (Hong Kong and Macau) Grant No.~12022516, and the RGC of Hong Kong SAR, China (GRF 16303621).
This work is supported in part by the JSPS KAKENHI grant Nos.~19H01895, 20H04727, 20H05853, and
24K00624.

\end{acknowledgments}

\appendix

\section{Quantization of a free field in an FLRW background}
\label{qffflrw}

This appendix reviews the quantization of a free field in an FLRW background.
In the main text, we evaluate the expectation values of the tensor field and the thermal radiation field.
The quantum state for the tensor fluctuation is chosen as the adiabatic vacuum during inflation.
We evolve the field operator from inflation to radiation dominant and then evaluate the expectation value with this quantum state.
We need to properly connect the mode functions during inflation and radiation dominant to find the correct primordial tensor spectrum during radiation dominant.
The Hamiltonian of the thermal radiation is time-dependent in the FLRW background, but the density operator is time-independent in the Heisenberg picture. 
We evaluate the Hamiltonian at the initial time of radiation dominant, i.e., at the reheating time, and then the density operator with this initial Hamiltonian.
Thus, we need to carefully select a state and the mode functions associated with the creation and annihilation operators we use.
This appendix provides detailed calculations for this procedure based on the fundamental matrix.
The results of this Appendix are used in Appendices~\ref{appthermal} and \ref{app:green}.

\subsection{Fundamental matrix}

Let us consider the Lagrangian of a massless scalar field in an FLRW background 
\begin{align}
	L_\chi [\chi, \chi'] = \frac{1}{2} \int d^3x  a^2[\chi'^2 -(\partial_i \chi)^2].
\end{align}
The conjugate momentum is defined as
\begin{align}
	\pi_\chi\equiv \frac{\delta L_\chi[\chi, \chi']}{\delta \chi'} = a^2\chi'.
\end{align}
The Hamiltonian is obtained as 
\begin{align}
	H_\chi[\chi, \pi_\chi] &= \int d^3x \pi_\chi \chi' - L_\chi[\chi, \chi'] 
	\notag 
	\\
	&= \frac{1}{2} \int d^3x  \left[\frac{\pi_\chi^2}{a^2} + a^2(\partial_i \chi)^2\right].
\end{align}
The dynamics of the field is given by the Hamilton equations
\begin{align}
	\frac{\delta H_\chi}{\delta \chi} &=-\pi_\chi' =  -a^2\partial^2 \chi,
	\\
	\frac{\delta H_\chi}{\delta \pi_\chi} & = \chi' = \frac{\pi_\chi}{a^2},
\end{align}
which, in Fourier space, can be expressed by the following matrix equation
\begin{align}
	Y_{\mathbf k}' = M_{\mathbf k} Y_{\mathbf k},\label{fundaeom}
\end{align}
where we defined
\begin{align}
	Y_{\mathbf k} = 
	\begin{pmatrix}
		\chi_{\mathbf k}
		\\
		\pi_{\chi, \mathbf k}
	\end{pmatrix}
	,
	M_{\mathbf k}
	=
	\begin{pmatrix}
		0 & \frac{1}{a^2} \\
		- a^2k^2 & 0
	\end{pmatrix}.
\end{align}
Meanwhile, $\chi$ can be either classical or quantum fields.
To solve the Hamilton equations, we introduce the fundamental matrix $\mathcal O_{\mathbf k}(\tau,\tau_0)$, such that 
\begin{align}
	\mathcal O_{\mathbf k}(\tau,\tau_0)' = M_{\mathbf k}\mathcal O_{\mathbf k}(\tau,\tau_0),~\mathcal O_{\mathbf k}(\tau_0,\tau_0)=
	\begin{pmatrix}
		1 & 0 
		\\
		0 & 1
	\end{pmatrix}
	.
\end{align}
Using the fundamental matrix, we find the solution for a given initial condition $Y_{\mathbf k}(\tau_0)$ as
\begin{align}
	Y_{\mathbf k}(\tau) = \mathcal O_{\mathbf k}(\tau,\tau_0) Y_{\mathbf k}(\tau_0).
\end{align}

In classical theory, we need an initial value of $Y_{\mathbf k}(\tau_0)$.
In quantum theory, $Y_{\mathbf k}(\tau_0)$ is composed of $q$-numbers, which we cannot specify a value at some time.
We promote the dynamical variables to operators in a Hilbert space and impose the simultaneous commutation relation for $\chi$ and $\pi_\chi$.
In Fourier space, the commutator is written as
\begin{align}
	[\hat \chi_{\mathbf k}, \hat \pi_{\chi, \mathbf k'}] = i(2\pi)^3\delta(\mathbf k+\mathbf k').\label{cq:com}
\end{align}
The Wronskian condition $\det |\mathcal O(\tau,\tau_0)|=1$ implies that we can impose \eqref{cq:com} at any time.
In contrast to the classical case, the commutation relation only fixes the normalization of the mode functions, and the ambiguity of the Bogoliubov transformation remains.
A quantum solution is fixed when we select a state to evaluate the operator.

Let us find the fundamental matrix.
By comparing the components of Eq.~\eqref{fundaeom}, one finds
\begin{align}
	-a^2 k^2 O_{{\mathbf k},11} &=(a^2	O_{{\mathbf k},11}')' ,
	\\
	-a^2 k^2 O_{{\mathbf k},12} &=(a^2	O_{{\mathbf k},12}')' .
\end{align}
We solve these differential equations with the initial conditions
\begin{align}
	O_{{\mathbf k},11}(\tau_0) &=1,~O_{{\mathbf k},11}'(\tau_0) = \frac{O_{{\mathbf k},21}(\tau_0)}{a(\tau_0)^2}=0 ,
	\\
	O_{{\mathbf k},12}(\tau_0) &=0,~O'_{{\mathbf k},12}(\tau_0) = \frac{O_{{\mathbf k},22}(\tau_0)}{a(\tau_0)^2}=  \frac{1}{a(\tau_0)^2}.
\end{align}
and then we find the rest of the components by computing
\begin{align}
	O_{{\mathbf k},21} &= a^2 O_{{\mathbf k},11}',\\
	O_{{\mathbf k},22} &= a^2 O_{{\mathbf k},12}'.
\end{align}

\subsection{Connection of mode functions}
Once we find fundamental matrixes, we can easily connect solutions in different regimes.
Let us consider fundamental matrixes for radiation dominant $\mathcal O_{\rm RD}$ and inflation $\mathcal O_{\rm dS}$ are found.
We consider that the inflationary phase stops at $\tau =-\tau_{\rm R}<0$ and then instantaneously transfers to radiation dominant at $\tau=\tau_{\rm R}>0$.
The scale factor and its derivative, i.e., the Hubble parameter, are continuous at the transition by defining
\begin{align}
	a(\tau) = 
	\begin{cases}
	&	-\frac{1}{H_{\rm inf.}\tau}, ~ \tau < -\tau_{\rm R} <0
		\\
	&	\frac{1}{H_{\rm inf.}\tau_{\rm R}}\left( \frac{\tau}{\tau_{\rm R}}\right), ~ 0 < \tau_{\rm R} < \tau,
	\end{cases}\label{scalefactora}
\end{align}
where $H_{\rm inf.}$ is the Hubble parameter at $\tau=\tau_{\rm R}$.
Then we find the field operator during radiation dominant as
\begin{align}
	\hat Y_{\mathbf k}(\tau) &= \mathcal O_{\rm RD,{\mathbf k}}(\tau,\tau_{\rm R}) \hat Y_{\mathbf k}(\tau_{\rm R})
	\notag 
	\\
	& = \mathcal O_{\rm RD,{\mathbf k}}(\tau,\tau_{\rm R})\mathcal O_{\rm dS,{\mathbf k}}(-\tau_{\rm R},\tau_{0}) \hat Y_{\mathbf k}(\tau_0).\label{connect:sol}
\end{align}
Creation and annihilation operators are obtained by a linear transformation of $\hat Y_{\mathbf k}(\tau_0)$ as
\begin{align}
	\binom{\hat d_{\mathbf k}}{\hat d^\dagger_{-\mathbf k}}  =\lim_{\tau_0\to -\infty}U_k(\tau_0) \hat Y_{\mathbf k}(\tau_0),\label{def:d}
\end{align}
in such a way that we find the spectral decomposition of the Hamiltonian:
\begin{align}
	\hat H_{\chi} = \int \frac{d^3 k}{(2\pi)^3} k \left( \hat d^\dagger_{\mathbf k}\hat d_{\mathbf k} + \frac{1}{2} [\hat d_{\mathbf k}, \hat d^\dagger_{\mathbf k}]  \right),\label{hem}
\end{align}
where we defined
\begin{align}
	U_k(\tau)	& \equiv 	
	\begin{pmatrix}
		a(\tau)\sqrt{\frac{k}{2}} & \frac{i}{a(\tau)\sqrt{2k}}
		\\
		a(\tau)\sqrt{\frac{k}{2}} & -\frac{i}{a(\tau)\sqrt{2k}}
	\end{pmatrix}.\label{defunitary}
\end{align}
Here we chose the adiabatic vacuum $|\psi \rangle$ as the initial condition by taking $\tau_0\to -\infty$:
\begin{align}
	d_{\mathbf k}|\psi \rangle = 0.
\end{align}
From Eq.~\eqref{cq:com} the creation and annihilation operators satisfy
\begin{align}
	[\hat d_{\mathbf k}, \hat d^\dagger_{-\mathbf k'}] = (2\pi)^3 \delta (\mathbf k+\mathbf k').
\end{align}
The inflationary mode functions for the $d$-operators are obtained as the expansion coefficients with respect to Eq.~\eqref{def:d}:
\begin{align}
	\begin{pmatrix}
		u^{\rm dS}_k(\tau) & u^{\rm dS *}_k(\tau)
		\\
		v^{\rm dS}_k(\tau) & v^{\rm dS *}_k(\tau)
	\end{pmatrix}
	= \lim_{\tau_0\to -\infty }   \mathcal O_{\rm dS}(\tau,\tau_{0})   U_k^{-1}(\tau_0).\label{mddefds}
\end{align}
Similarly, we may define creation and annihilation operators for the vacuum state $|0\rangle$ at the initial time of radiation dominant:
\begin{align}
	\binom{\hat b_{\mathbf k}}{\hat b^\dagger_{-\mathbf k}}  =U_k(\tau_{\rm R}) \hat Y_{\mathbf k}(\tau_{\rm R}),\label{bdef}
\end{align}
with
\begin{align}
	b_{\mathbf k}|0 \rangle = 0.
\end{align}
Eq.~\eqref{cq:com} also yields
\begin{align}
	[\hat b_{\mathbf k}, \hat b^\dagger_{-\mathbf k'}] = (2\pi)^3 \delta (\mathbf k+\mathbf k').
\end{align}
Mode functions associated with the $b$-operators are
\begin{align}
	\begin{pmatrix}
		u^{\rm RD}_k(\tau) & {u^{\rm RD}}^*_k(\tau)
		\\
		v^{\rm RD}_k(\tau) & {v^{\rm RD}}^*_k(\tau)
	\end{pmatrix}
	= \mathcal O_{\rm RD}(\tau,\tau_{\rm R})U_k^{-1}(\tau_{\rm R}).\label{mddefrd}
\end{align}
$b$-operators will be used for quantization of thermal radiation in Appendix~\ref{appthermal} as an initial thermal distribution is defined with respect to $|0\rangle$.
For primordial gravitational waves during radiation dominance, we may expand the field operators not by the $b$-operators but the $d$-operators.
From Eq.~\eqref{connect:sol}, the mode functions during radiation dominant with respect to the $d$-operators are written as
\begin{align}
	&\begin{pmatrix}
		u^{\rm dS \to RD}_k(\tau) & u^{\rm dS \to RD*}_k(\tau)
		\\
		v^{\rm dS \to RD}_k(\tau) & v^{\rm dS \to RD*}_k(\tau)
	\end{pmatrix}
	\notag
	\\
	&= \lim_{\tau_0\to -\infty }    \mathcal O_{\rm RD}(\tau,\tau_{\rm R})\mathcal O_{\rm dS}(-\tau_{\rm R},\tau_{0})   U_k^{-1}(\tau_0).\label{dstordmd}
\end{align}
Thus, we can write the mode functions systematically once we find the fundamental matrix and fix the unitary operator $U_k(\tau)$.
In the following subsections, we obtain the concrete form of the fundamental matrixes during inflation and radiation dominant.

In a more realistic scenario, we may consider inflation stops at some time $\tau = \tau_{\rm e}<0$, and the early matter dominant is realized.
Then reheating happens, and the universe is dominated by radiation.
In this case, we need to find the fundamental matrix for the intermediate stage, $\mathcal O_{\rm eMD}(\tau_{\rm R},-\tau_{\rm e})$.
Then, the mode functions are found as 
\begin{align}
	&\begin{pmatrix}
		u^{\rm dS \to eMD \to RD}_k(\tau) & u^{\rm dS \to eMD \to RD*}_k(\tau)
		\\
		v^{\rm dS \to eMD \to RD}_k(\tau) & v^{\rm dS \to eMD \to RD*}_k(\tau)
	\end{pmatrix}
	\notag
	\\
	&= \lim_{\tau_0\to -\infty }    \mathcal O_{\rm RD}(\tau,\tau_{\rm R})\mathcal O_{\rm eMD}(\tau_{\rm R},-\tau_{\rm e}) \mathcal O_{\rm dS}(\tau_{\rm e},\tau_{0})   U_k^{-1}(\tau_0).\label{dstordmdemd}
\end{align}
As the transition from inflation to radiation is not instantaneous in this case, the Hubble parameter at reheating time differs from the inflationary Hubble parameter, while we identified those in the main text.

\subsection{Fundamental matrixes in de Sitter and radiation dominant Universe}
In de Sitter spacetime, the scale factor is given by the first line of Eq.~\eqref{scalefactora}.
In this case, we find
\begin{align}
	O^{\rm dS}_{11}(\tau,\tau_0) &=\frac{\tau  \cos (k (\tau -\tau_0))}{\tau_0}-\frac{\sin (k (\tau -\tau_0))}{k \tau_0},
	\\
	O^{\rm dS}_{12}(\tau,\tau_0) &=\frac{H_{\rm inf.}^2 (\tau_0-\tau ) \cos (k (\tau -\tau_0))}{k^2}
	\notag
	\\
	&+\frac{H_{\rm inf.}^2 \left(k^2 \tau  \tau_0+1\right) \sin (k (\tau -\tau_0))}{k^3},
	\\
	O^{\rm dS}_{21}(\tau,\tau_0) &=-\frac{k \sin (k (\tau -\tau_0))}{H_{\rm inf.}^2 \tau  \tau_0},
	\\
	O^{\rm dS}_{22}(\tau,\tau_0) &=\frac{\sin (k (\tau -\tau_0))}{k \tau }+\frac{\tau_0 \cos (k (\tau -\tau_0))}{\tau }.
\end{align}
Then, the mode functions \eqref{mddefds} are
\begin{align}
	\lim_{\tau_0\to -\infty}u^{\rm dS}_{q}(\tau) &= \frac{iH_{\rm inf.} e^{-i q (\tau -\tau_0) } (1+iq \tau )}{\sqrt{2} q^{3/2}},
	\\
	\lim_{\tau_0\to -\infty}{v^{\rm dS}_{q}}(\tau) &=\frac{i \sqrt{q} e^{-i q (\tau -\tau_0) }}{\sqrt{2} H_{\rm inf.} \tau }.
\end{align}
In radiation dominant, the scale factor is given by the second line of Eq.~\eqref{scalefactora}.
With this scale factor, the fundamental matrix is found as
\begin{align}
	O^{\rm RD}_{11}(\tau,\tau_{\rm R}) &=\frac{\sin (k (\tau -\tau_{\rm R}))}{k \tau }+\frac{\tau_{\rm R} \cos (k (\tau -\tau_{\rm R}))}{\tau },
	\\
	O^{\rm RD}_{12}(\tau,\tau_{\rm R}) &=\frac{H_{\rm inf.}^2 \tau_{\rm R}^3 \sin (k (\tau -\tau_{\rm R}))}{k \tau },
	\\
	O^{\rm RD}_{21}(\tau,\tau_{\rm R}) &=\frac{\left(k^2 (-\tau ) \tau_{\rm R}-1\right) \sin (k (\tau -\tau_{\rm R}))}{H_{\rm inf.}^2 k \tau_{\rm R}^4}
	\notag 
	\\
	&+\frac{(\tau -\tau_{\rm R}) \cos (k (\tau -\tau_{\rm R}))}{H_{\rm inf.}^2 \tau_{\rm R}^4},
	\\
	O^{\rm RD}_{22}(\tau,\tau_{\rm R}) &=\frac{\tau  \cos (k (\tau -\tau_{\rm R}))}{\tau_{\rm R}}-\frac{\sin (k (\tau -\tau_{\rm R}))}{k \tau_{\rm R}}.
\end{align}
Then, the mode functions \eqref{mddefrd} are
\begin{align}
	 u^{\rm RD}_q(\tau) &= \frac{H_{\rm inf.} \tau_{\rm R} \left(\sin (q (\tau -\tau_{\rm R}))+q \tau_{\rm R} e^{-i q (\tau -\tau_{\rm R})}\right)}{\sqrt{2} q^{3/2} \tau },\label{md_u:rd}
	\\
	 v^{\rm RD}_q(\tau) &= \frac{(-i q \tau  \tau_{\rm R}+\tau -\tau_{\rm R}) \cos (q (\tau -\tau_{\rm R}))}{\sqrt{2} H_{\rm inf.} q^{1/2} \tau_{\rm R}^3} 
	 \notag 
	 \\
	 &-\frac{(1+q \tau_{\rm R} (q \tau -i)) \sin (q (\tau -\tau_{\rm R}))}{\sqrt{2} H_{\rm inf.} q^{3/2} \tau_{\rm R}^3}.\label{md_v:rd}
\end{align}

\subsection{Primordial tensor power spectrum}
\label{apppgw}

In the main text, we evaluated the tensor power spectrum in the radiation dominant.
The mode functions during radiation dominant are expanded with respect to the creation and annihilation operators for the inflationary adiabatic vacuum.
This section provides the correct mode functions using the abovementioned fundamental matrix.

For the tensor fluctuations, the mode functions have additional normalization factors.
We find
\begin{align}
	u_{h,q}(\tau) &= \frac{2}{M_{\rm pl}} u_{q}(\tau),\label{uh}
	\\
	v_{h,q}(\tau) &= \frac{M_{\rm pl}}{2} v_{q}(\tau).\label{vh}
\end{align}
The linear tensor mode during inflation is expressed as
\begin{align}
		 \hat h_i{}^j(\tau,\mathbf q) &= \sum_{s=\pm 2} e^{(s)}_{i}{}^j(\mathbf q)
	 \left[u^{\rm dS}_{h,q}(\tau) \hat d^{(s)}_{\mathbf q} + u^{\rm dS*}_{h,q}(\tau) \hat d^{(s)\dagger}_{-\mathbf q}\right],
\end{align}
where the creation and annihilation operators satisfy
\begin{align}
	[\hat d^{(s)}_{\mathbf q},\hat d^{(s')\dagger}_{-\mathbf q'}]=\delta^{ss'}(2\pi)^3 \delta(\mathbf q+\mathbf q'),
\end{align}
and the polarization tensors are normalized as
\begin{align}
	e^{(s)}_{i}{}^j(\mathbf k)(e^{(s')}_j{}^i(\mathbf k))^*  = \delta^{ss'}.\label{def:nom:pol}
\end{align}
The annihilation operator defines the adiabatic vacuum
\begin{align}
	d^{(s)}_{\mathbf k}|\psi \rangle = 0.
\end{align}
The density operator for the tensor fluctuation in the adiabatic vacuum during inflation is
\begin{align}
	\hat \varrho_h \equiv |\psi\rangle \langle \psi|.\label{denshdef}
\end{align}
The equal time power spectrum is then evaluated as
\begin{align}
	&{\rm Tr}\left[  \hat \varrho_h  \hat h_i{}^j(\tau,\mathbf q)\hat h_j{}^i(\tau,\mathbf q') \right]
	\notag 
	\\
	&=	|u^{\rm dS}_{h,q}(\tau)|^2\sum_{s=\pm 2} \langle \psi | \hat d^{(s)}_{\mathbf q} \hat d^{(s)\dagger}_{-\mathbf q'} |\psi \rangle
	\notag
	\\
	&=	2|u^{\rm dS}_{h,q}(\tau)|^2  (2\pi)^3 \delta(\mathbf q+\mathbf q')
	.
\end{align}
Thus, the power spectrum during inflation is obtained as
\begin{align}
	P^{\rm dS}_h(\tau, q) =2|u^{\rm dS}_{h,q}(\tau)|^2 = \frac{4H_{\rm inf.}^2}{q^3M^2_{\rm pl}}(1+q^2\tau^2).
\end{align}
During radiation dominant, the linear tensor mode with respect to the $d$-operators is
\begin{align}
&		 \hat h_i{}^j(\tau,\mathbf q) =\sum_{s=\pm 2} e^{(s)}_{i}{}^j(\mathbf q)
\notag 
\\
\times & 
	 \left[u^{\rm dS \to RD}_{h,q}(\tau) \hat d^{(s)}_{\mathbf q} + u^{\rm dS \to RD*}_{h,q}(\tau) \hat d^{(s)\dagger}_{-\mathbf q}\right].
\end{align}
We immediately find these mode functions from Eq.~\eqref{dstordmd}.
In $\tau_0\to -\infty$ limit, we find
\begin{align}
	\lim_{\tau_0\to -\infty} u^{\rm dS \to RD}_{h,q}(\tau) &=\frac{2}{M_{\rm pl}} \frac{ie^{i q \tau_0} H_{\rm inf.}  \sin (q \tau )}{\sqrt{2} q^{5/2} \tau },\label{udsdrlim}
	\\
	\lim_{\tau_0\to -\infty} v^{\rm dS \to RD}_{h,q}(\tau) & =\frac{M_{\rm pl}}{2} ie^{i q \tau_0} \frac{ q\tau \cos(q\tau) -   \sin (q \tau )}{\sqrt{2} H_{\rm inf.} q^{5/2} \tau_{\rm R}^4 }.
\end{align}
Note that Eqs.~\eqref{mddefrd} does not give these mode functions but \eqref{dstordmd} since we expand the field operator by $d$-operators.
Then, we find the linear tensor power spectrum in the radiation-dominated universe:
\begin{align}
	P^{\rm RD}_h(\tau, q) =2 |u^{\rm dS \to RD}_{h,q}(\tau)|^2= \frac{4H_{\rm inf.}^2}{q^3M_{\rm pl}^2} \frac{\sin^2(q\tau)}{q^2\tau^2}.
\end{align}
Note that we find the same result in the classical cosmological perturbation theory with the super horizon initial condition set by inflation.

\section{Thermal field theory}
\label{appthermal}

In this appendix, we derive the thermal average of a massless scalar field in the unperturbed and perturbed FLRW spacetime.
We evaluate the Hamiltonian at the initial time of radiation dominant and define the density operator to calculate the thermal average of energy density.

\subsection{Unperturbed FLRW spacetime}

Let us first consider a scalar field in an FLRW spacetime, which is discussed in appendix~\ref{qffflrw}.
For a given unitary operator Eq.~\eqref{defunitary}, we find the time-dependent Hamiltonian~\eqref{hem}.
The second term of the free Hamiltonian is the constant shift of the vacuum energy.
The time dependence of the first term arises because we consider an expanding background.
Given a Hamiltonian operator, a thermal state of the inverse temperature $\beta$ is characterized by the density operator
\begin{align}
	\hat \varrho_{\chi} \equiv \frac{e^{-\beta \hat H_{\chi }}}{Z},~Z\equiv {\rm Tr}[e^{-\beta \hat H_{\chi }}].\label{defrhoapp}
\end{align}
We may ignore the constant shift arising from the vacuum energy, as it is canceled by the normalization factor in Eq.~\eqref{defrhoapp}.

It is convenient to find a matrix representation of the density operator with respect to the $n$-particle states.
First, let us write the density operator as a product of those for each $k$-mode:
\begin{align}
	 \hat \varrho_{\chi}  =  \prod_k \hat \varrho_k,
\end{align}
where we defined
\begin{align}
	\hat \varrho_k \equiv \frac{e^{-\beta  k  \hat B^{\dagger}_{k}  \hat B_{k}}}{Z_k},~Z_k \equiv {\rm Tr}\left[ e^{-\beta  k  \hat B^{\dagger}_{k}  \hat B_{k}} \right]_k, 
\end{align}
Note that we defined the dimensionless creation and annihilation operators as $B \equiv b/\sqrt{V}$ with $V \equiv [b_{\mathbf k},b^\dagger_{\mathbf k}]$.
$k$ is a discrete index specifying $\mathbf k$ in this context.
${\rm Tr}[ \cdots ]_k$ is the partial trace for the index $k$.
The product of $k$ is understood as
\begin{align}
	\ln \left[ \prod_k \hat \varrho_k \right] = \sum_{k} \ln \hat \varrho_k,~\sum_k \equiv  V \int \frac{d^3 k }{(2\pi)^3}.
\end{align}
In this convention, the 3-dimensional delta function is written as
\begin{align}
	(2\pi)^3 \delta(\mathbf p-\mathbf k) = V \delta_{kp}. 
\end{align}
The $n$-particle states can be generated as usual by using the creation operator $\hat B^\dagger_{k}$ and the vacuum state $|0\rangle$ such that $\hat B_{k}|0\rangle =0$:
\begin{align}
	|n_{k}\rangle &= \frac{(\hat B^\dagger_{k})^n |0\rangle }{\sqrt{n_{k}! }},
\end{align}
which yields
\begin{align}
		\langle n_{k} | m_{p} \rangle &= \delta_{nm}\delta_{kp},
	\\
	 \hat B^\dagger_{k}\hat B_{k} | n_{k} \rangle &= n_{k}| n_{k} \rangle,
	 \\
	 \hat B_{k} \hat B^\dagger_{k} | n_{k} \rangle &= (n_{k}+1 )| n_{k} \rangle  .
\end{align}
Using the spectral decomposition of 1 for the $k$ mode, we find the matrix representation: 
\begin{align}
	\hat \varrho_k &= \sum_{m_k}|m_k \rangle \langle m_k| \hat \varrho_k \sum_{n_k}|n_k \rangle \langle n_k|
	\notag \\
	&= \frac{1}{Z_k}\sum_{m_k}\sum_{n_k} |m_k \rangle \langle m_k| e^{-\beta  k  \hat B^{\dagger}_{k}  \hat B_{k}  } |n_k \rangle \langle n_k|
		\notag \\
	&= \frac{1}{Z_k}\sum_{m_k}\sum_{n_k} e^{-\beta  k n_k} |m_k \rangle \langle m_k  |n_k \rangle \langle n_k|
			\notag \\
	&= \frac{1}{Z_k} \sum_{n_k} e^{-\beta  k n_k}   |n_k \rangle \langle n_k|.
\end{align}
In the Heisenberg picture, the state is fixed initially, and we consider the evolution of physical operators.
Therefore, hereafter the density operator~\eqref{defrhoapp}, i.e., $B$ and $B^\dagger$ are fixed at the initial time of radiation dominant, $\tau =\tau_{\rm R}$.
Using the creation and annihilation operators defined at the initial time, the field operator at a later time is expanded into
\begin{align}
	\hat \chi(\tau,\mathbf x) =
	\frac{\sum_{k}}{\sqrt{V}}e^{i\mathbf k\cdot \mathbf x}\left[u^{\rm RD}_{k}(\tau) \hat B_{k} + u^{\rm RD*}_{k}(\tau) \hat B^\dagger_{-k}\right]
	.
\end{align}
$u^{\rm RD}_{k}$is the positive frequency mode function defined in Eq.~\eqref{md_u:rd}.
Since ${\rm Tr}[\hat \varrho_{\chi} \hat B]=0$, one finds 
\begin{align}
		&{\rm Tr}\left[ \hat \varrho_{\chi}  \hat B^\dagger_{-p_1} \hat B_{p_2} \right]
		\notag 
		\\
		=&\delta_{-p_1 p_2}	{\rm Tr}\left[ \hat \varrho_{p_2}  \hat B^\dagger_{p_2} \hat B_{p_2} \right] 
		\notag 
		\\=&
		 \delta_{-p_1 p_2}	\frac{1}{Z_{p_2}}\sum_{n_{p_2}} e^{-\beta p_2 n_{p_2}} \langle n_{p_2}|  \hat B^\dagger_{p_2} \hat B_{p_2} |n_{p_2} \rangle 
		\notag 
		\\=&  
		\delta_{-p_1 p_2}	\frac{1}{Z_{p_2}}\sum_{n_{p_2}} e^{-\beta p_2 n_{p_2}} n_{p_2}
		\notag 
		\\
		=&
		\delta_{-p_1 p_2}	\frac{1}{ e^{\beta p_2 } -1 }.
\end{align}
Then we get
\begin{align}
	&{\rm Tr}\left[ \hat \varrho_{\chi}  \hat \chi(x_1)\hat \chi(x_2) \right] 	= \int \frac{d^3k_1 }{(2\pi)^3}e^{i\mathbf k_1\cdot ( \mathbf x_1- \mathbf x_2 ) }
	\notag 
	\\
	&\times \left[
		u^{\rm RD}_{k_1}(\tau_1)  u^{\rm RD*}_{k_1}(\tau_2) (f_{\beta k_1}+1)
	 + u^{\rm RD*}_{k_1}(\tau_1) u^{\rm RD}_{k_1}(\tau_2)  f_{\beta k_1}
	 \right],\label{chichif}
\end{align}
where the Planck distribution is defined as 
\begin{align}
	f_{\beta k} \equiv \frac{1}{ e^{\beta k}-1 }.	
\end{align}
Using Eq.~\eqref{chichif} for $x_1=x_2$, one finds
\begin{align}
	\frac{a^2}{2} {\rm Tr}\left[\hat \varrho_{\chi} \partial_i\hat \chi\partial_j\hat \chi	\right]&=  \frac{a^2\delta_{ij}}{6}\int \frac{ d^3 p_1}{(2\pi)^3} 2 p_1^2 f_{{\beta p_1}}   |u^{\rm RD}_{ p_1}|^2
\notag 
\\
&+  \frac{a^2\delta_{ij}}{6} \int \frac{ d^3 p_1}{(2\pi)^3}p_1^2  |u^{\rm RD}_{ p_1}|^2,\label{eqijchichi}
\\
\frac{1}{2a^2}{\rm Tr}\left[\hat \varrho_{\chi} \hat \pi^2_{\chi} \right]&=  \frac{1}{2a^2}\int \frac{ d^3 p_1}{(2\pi)^3}  2 f_{\beta p_1}    |v^{\rm RD}_{ p_1}|^2
\notag 
\\
&
+  \frac{1}{2a^2} \int \frac{ d^3 p_1}{(2\pi)^3}  |v^{\rm RD}_{ p_1}|^2. 
\end{align}
The first terms are the finite temperature contribution, and the second is divergent vacuum energy.
Substituting Eqs.~\eqref{md_u:rd} and \eqref{md_v:rd} into these equations and integrating the momentum in the high-temperature limit, we find the asymptotic formulas
\begin{align}
	\lim_{\beta\to 0}a^2 {\rm Tr}\left[\hat \varrho_{\chi} \partial_i\hat \chi\partial_j\hat \chi	\right] & = \frac{\pi^2}{90 \beta^4} \delta_{ij} ,\label{gradav}
	 \\
	\lim_{\beta\to 0} a^{-2} {\rm Tr}\left[\hat \varrho_{\chi} \hat \pi^2_{\chi} \right]& = \frac{\pi^2}{30  \beta^4} 
	 .\label{dotav}
\end{align}
The energy density and pressure are divided by the volume element, $\sqrt{-g} = a^4$ in the end.
The subdominant contributions are, e.g.,  particle production in the radiation background, which dilutes faster than the dominant part. 
We ignored the zero temperature part, and then we found the equation of state for radiation.
The renormalization prescription for the zero temperature part can be found in, e.g., Refs.~\cite{Landsman:1986uw}.
Eqs.~\eqref{gradav} and \eqref{dotav} can also be found by taking $q\to \infty$ limit for the mode functions a priori.
In the high-frequency limit of the radiation particle, we find
\begin{align}
	\lim_{q\to \infty} a^2|u^{\rm RD}_{q}|^2 & = \frac{1}{2 q} ,
	\\
	\lim_{q\to \infty} a^{-2}|v^{\rm RD}_{q}|^2 & = \frac{q}{2},
\end{align}
and then we find the same energy density and pressure.

\subsection{Perturbed FLRW}

Next, let us consider a scalar field minimally coupled to gravity.
In this case, we consider a local thermal state density operator $\hat \varrho_\chi$.
From the equivalence principle, we can take local FLRW frames.
Gravitational waves are eliminated in this frame and the thermal average is defined there.
When taking the average in a local frame, the size of the FLRW region matters.
If the typical wavelength of the radiation field $\lambda_\chi\sim \beta^{-1}$ or the mean free path of $\chi$ is sufficiently smaller than the curvature radius, the thermal average in the FLRW frame is approximated by the global one.
To be more specific, the thermal average in the local FLRW frame should be given as
\begin{align}
	{\rm Tr}[\hat \varrho_{\chi } \partial_{(i)}\hat \chi\partial_{(j)}\hat \chi ] =  a^2 P_\gamma   \delta_{(i)(j)} + \mathcal O\left(R_{(i)(j)} \lambda_\chi^2\right),  
\end{align}
where an index in a bracket implies the local FLRW coordinate.
Hence, using the tetrad $E_{\mu}{}^{(\alpha)}$, the local thermal average of the free field in the global FLRW frame is
\begin{align}
    {\rm Tr}[\hat \varrho_{\chi } \partial_{\mu}\hat \chi\partial_{\nu}\hat \chi ] = 
    E_\mu{}^{(\alpha)}
    E_\mu{}^{(\beta)}
    {\rm Tr}[\hat \varrho_{\chi} \partial_{(\alpha)}\hat \chi\partial_{(\beta)}\hat \chi ].\label{deflocav}
\end{align}
Then, we find
\begin{align}
	{\rm Tr}[\hat \varrho_{\chi } \partial_{i}\hat \chi\partial_{j}\hat \chi ]= P g_{ij} + \mathcal O\left(E_{i}{}^{(k)}E_{j}{}^{(l)} R_{(k)(l)} \lambda_\chi^2\right).\label{Pij}
\end{align}
The Ricci tensor is $\mathcal O(\partial^2 h)$, so we have
\begin{align}
	E_{i}{}^{(k)}E_{j}{}^{(l)} R_{(k)(l)} \lambda_\chi^2= \mathcal O( h \lambda_\chi^2/\lambda_h^2).
\end{align}
We want to determine Eq.~\eqref{Pij} to an accuracy of $\mathcal O(h)$.
This is possible when $\lambda_\chi/\lambda_h \ll 1$.

\subsection{Wick's theorem at finite temperature}

We used Wick's theorem at finite temperature in the main text.
This section derives the theorem.
In a general case, the thermal average of an operator is written as
\begin{align}
	{\rm Tr} \left[ \hat \varrho  \mathcal O_{p_1,\cdots ,p_n} \right] 
	&=  {\rm Tr} \left[  \left(\prod_{k\in \{p_n\}} \hat \varrho_k \right) \mathcal O_{p_1,\cdots ,p_n} \right], 
\end{align}
where $\{p_n\}$ implies possible permutations of $\{p_1,\cdots ,p_n\}$. 
For example, for an operator with four momenta, we find
\begin{align}
	&	{\rm Tr} \left[ \hat \varrho  \mathcal O_{p_1,p_2,p_3,p_4} \right] 
	\notag 
	\\
	&={\rm Tr}\left[ \hat \varrho_{p_1} \mathcal O_{p_1,p_1,p_1,p_1} \right]_{p_1=p_2=p_3=p_4} 
	\notag 
	\\
	& +
	{\rm Tr}\left[ \hat \varrho_{p_1}\hat \varrho_{p_2} \mathcal O_{p_1,p_1,p_1,p_2} \right]_{p_1=p_2=p_3\neq p_4} + 3~{\rm perms}.
	\notag 
	\\
	& +
	{\rm Tr}\left[ \hat \varrho_{p_1}\hat \varrho_{p_2} \mathcal O_{p_1,p_1,p_2,p_2} \right]_{p_1=p_2\neq p_3=p_4} + 2~{\rm perms}.
	\notag 
	\\
	& +
	{\rm Tr}\left[ \hat \varrho_{p_1}\hat \varrho_{p_2}\hat \varrho_{p_3} \mathcal O_{p_1,p_2,p_3,p_1} \right]_{p_1=p_4\neq p_2\neq p_3} + 5~{\rm perms}  
	\notag 
	\\
	&+ {\rm Tr}\left[ \hat \varrho_{p_1}\hat \varrho_{p_2}\hat \varrho_{p_3}\hat \varrho_{p_4} \mathcal O_{p_1,p_1,p_3,p_4} \right]_{p_1\neq p_4\neq p_2\neq p_3}.\label{generalexp}
\end{align}
Wick's theorem at finite temperature is more complicated than the one at zero temperature in the presence of the density operator.
For simplicity, let us consider 
\begin{align}
	{\rm Tr}\left[ \hat \varrho_{\chi}  \hat \chi(x_1)\hat \chi(x_2)\hat \chi(x_3)\hat \chi(x_4) \right],
\end{align}
following the expansion theorem~\eqref{generalexp}.

\begin{widetext}	
After straightforward calculation, we find
\begin{align}
		&{\rm Tr}\left[ \hat \varrho_{\chi}  \hat \chi(x_1)\hat \chi(x_2)\hat \chi(x_3)\hat \chi(x_4) \right] 
		\notag 
		\\
		&= \frac{1}{V^2} \sum_{k_1k_2k_3k_4}
		e^{i\mathbf k_1\cdot \mathbf x_1}
		e^{i\mathbf k_2\cdot \mathbf x_2}
		e^{i\mathbf k_3\cdot \mathbf x_3}
		e^{i\mathbf k_4\cdot \mathbf x_4}
		\notag 
		\\
		&\times 
		{\rm Tr}\bigg[ \hat \varrho
		 u^{\rm RD}_{ k_1}(\tau_1) u^{\rm RD}_{ k_2}(\tau_2) u^{\rm RD*}_{ k_3}(\tau_3) u^{\rm RD*}_{ k_4}(\tau_4) \hat B_{k_1}	 \hat B_{k_2}		 \hat B^\dagger_{-k_3}	 \hat B^\dagger_{-k_4}
			+ \hat \varrho u^{\rm RD}_{ k_1}(\tau_1) u^{\rm RD*}_{ k_2}(\tau_2)  u^{\rm RD}_{ k_3}(\tau_3) u^{\rm RD*}_{ k_4}(\tau_4) \hat B_{k_1}  \hat B^\dagger_{-k_2}	 \hat B_{k_3}  \hat B^\dagger_{-k_4}
		\notag 
		\\
		&	+ \hat \varrho u^{\rm RD}_{ k_1}(\tau_1)u^{\rm RD*}_{ k_2}(\tau_2)u^{\rm RD*}_{ k_3}(\tau_3)u^{\rm RD}_{ k_4}(\tau_4)  \hat B_{k_1}  \hat B^\dagger_{-k_2}  \hat B^\dagger_{-k_3} \hat B_{k_4}
		 + \hat \varrho u^{\rm RD*}_{ k_1}(\tau_1) u^{\rm RD*}_{ k_2}(\tau_2) u^{\rm RD}_{ k_3}(\tau_3) u^{\rm RD}_{ k_4}(\tau_4) \hat B^\dagger_{-k_1} \hat B^\dagger_{-k_2} \hat B_{k_3} \hat B_{k_4}
		\notag 
		\\
		& + \hat \varrho u^{\rm RD*}_{ k_1}(\tau_1) u^{\rm RD}_{ k_2}(\tau_2) u^{\rm RD}_{ k_3}(\tau_3) u^{\rm RD*}_{ k_4}(\tau_4) \hat B^\dagger_{-k_1} \hat B_{k_2} \hat B_{k_3} \hat B^\dagger_{-k_4}
		 + \hat \varrho u^{\rm RD*}_{ k_1}(\tau_1) u^{\rm RD}_{ k_2}(\tau_2) u^{\rm RD*}_{ k_3}(\tau_3) u^{\rm RD}_{ k_4}(\tau_4) \hat B^\dagger_{-k_1} \hat B_{k_2} \hat B^\dagger_{-k_3} \hat B_{k_4}
	\bigg].
\end{align}
The last term can be calculated as follows:
\begin{align}
	&\frac{1}{V^2} \sum_{k_1k_2k_3k_4}
		e^{i\mathbf k_1\cdot \mathbf x_1}
		e^{i\mathbf k_2\cdot \mathbf x_2}
		e^{i\mathbf k_3\cdot \mathbf x_3}
		e^{i\mathbf k_4\cdot \mathbf x_4} 
		u^{\rm RD*}_{ k_1}(\tau_1) u^{\rm RD}_{ k_2}(\tau_2) u^{\rm RD*}_{ k_3}(\tau_3) u^{\rm RD}_{ k_4}(\tau_4) 
		{\rm Tr}\bigg[ \hat \varrho \hat B^\dagger_{-k_1} \hat B_{k_2} \hat B^\dagger_{-k_3} \hat B_{k_4} \bigg] 
	\notag 
	\\
	&=\frac{1}{V^2} \sum_{k_1}
		e^{i\mathbf k_1\cdot ( \mathbf x_1 - \mathbf x_2 + \mathbf x_3 - \mathbf x_4 ) }
		u^{\rm RD*}_{ k_1}(\tau_1) u^{\rm RD}_{ k_1}(\tau_2) u^{\rm RD*}_{ k_1}(\tau_3) u^{\rm RD}_{ k_1}(\tau_4) \sum_{n_{k_1}} \frac{e^{-\beta k_1 n_{k_1}}}{Z_{k_1}} n_{k_1}^2 
	\notag 
	\\
	& + \frac{1}{V^2} \sum_{k_1k_3}
		e^{i\mathbf k_1\cdot (\mathbf x_1 - \mathbf x_2) }
		e^{i\mathbf k_3\cdot (\mathbf x_3 - \mathbf x_4) }
		u^{\rm RD*}_{ k_1}(\tau_1) u^{\rm RD}_{ k_1}(\tau_2) u^{\rm RD*}_{ k_3}(\tau_3) u^{\rm RD}_{ k_3}(\tau_4)\sum_{n_{k_1}} \sum_{n_{k_3}} \frac{e^{-\beta k_1 n_{k_1}}}{Z_{k_1}}\frac{e^{-\beta k_3 n_{k_3}}}{Z_{k_3}}  n_{k_1}n_{k_3} \Big|_{k_1\neq k_3}
		\notag 
	\\
	& + \frac{1}{V^2} \sum_{k_1k_3}
		e^{i\mathbf k_1\cdot ( \mathbf x_1 - \mathbf x_4)  }
				e^{i\mathbf k_3\cdot ( \mathbf x_3 - \mathbf x_3 ) }
				u^{\rm RD*}_{ k_1}(\tau_1) u^{\rm RD}_{ k_3}(\tau_2) u^{\rm RD*}_{ k_3}(\tau_3) u^{\rm RD}_{ k_1}(\tau_4) \sum_{n_{k_1}} \sum_{n_{k_3}} \frac{e^{-\beta k_1 n_{k_1}}}{Z_{k_1}}\frac{e^{-\beta k_3 n_{k_3}}}{Z_{k_3}}  n_{k_1}(n_{k_3}+1) \Big|_{k_1\neq k_3}.
\end{align}
Now, we can see 
\begin{align}
	\frac{1}{V^2} \sum_{k_1} = \frac{1}{V} \int \frac{d^3 k_1}{(2\pi)^3} \ll  \frac{1}{V^2} \sum_{k_1k_3} = \int \frac{d^3 k_1d^3 k_3 }{(2\pi)^6}.
\end{align}
With this hierarchy, we may split the 4-point function into the product of 2-point functions:
\begin{align}
	{\rm Tr}\left[ \hat \varrho_{\chi}  \hat \chi(x_1)\hat \chi(x_2)\hat \chi(x_3)\hat \chi(x_4) \right] 
	&=  
	{\rm Tr}\left[ \hat \varrho_{\chi}  \hat \chi(x_1)\hat \chi(x_2) \right]{\rm Tr}\left[ \hat \varrho_{\chi}  \hat \chi(x_3)\hat \chi(x_4) \right]
	+
	{\rm Tr}\left[ \hat \varrho_{\chi}  \hat \chi(x_1)\hat \chi(x_3) \right]{\rm Tr}\left[ \hat \varrho_{\chi}  \hat \chi(x_2)\hat \chi(x_4) \right]
	\notag 
	\\
	&+
	{\rm Tr}\left[ \hat \varrho_{\chi}  \hat \chi(x_1)\hat \chi(x_4) \right]{\rm Tr}\left[ \hat \varrho_{\chi}  \hat \chi(x_2)\hat \chi(x_3) \right].\label{wick:finite}
\end{align}
\end{widetext}

\section{Green functions}
\label{app:green}

This appendix summarizes concrete expressions of various Green functions used in the main text.
The scalar and tensor retarded Green functions are defined as
\begin{align}
	G(x,y)&\equiv i a(y^0)^2  [ \hat \chi(x) , \hat \chi (y)]\theta(x^0-y^0),\label{rt:chi}
	\\
	G_{ij}{}^{kl}(x,y)& \equiv  \frac{M_{\rm pl}^2}{4} i a(y^0)^2  [ \hat h_{ij}(x) , \hat h^{kl}(y)]\theta(x^0-y^0),\label{rt:h}
\end{align}
and the Keldysh Green function is defined as
\begin{align}
		G^K(y,z) &\equiv  {\rm Tr}\left[ \hat \varrho_\chi \frac{ \hat  \chi(z)  \hat \chi(y) 
	 + \hat  \chi(y)  \hat \chi(z)  
	}{2}\right].\label{kd:chi}
\end{align}
The Fourier integrals of the scalar field and the tensor fluctuation during radiation dominant are
\begin{align}
	\hat \chi(\tau,\mathbf x) &= \int \frac{d^3 k}{(2\pi)^3} e^{i\mathbf k \cdot \mathbf x} 
	\left[u^{\rm RD}_{k}(\tau) \hat b^{(0)}_{\mathbf k} + u^{\rm RD*}_{k}(\tau) \hat b^{(0)\dagger}_{-\mathbf k}\right],
	\\
	 \hat h_i{}^j(\tau,\mathbf x) &= \sum_{s=\pm 2}\int \frac{d^3 k}{(2\pi)^3} e^{i\mathbf k \cdot \mathbf x} e^{(s)}_{i}{}^j(\mathbf k)
	 \notag 
	 \\
	 &\times 
	 \left[u^{\rm RD}_{h,k}(\tau) \hat b^{(s)}_{\mathbf k} + u^{\rm RD*}_{h,k}(\tau) \hat b^{(s)\dagger}_{-\mathbf k}\right],
\end{align}
with the commutation relation
\begin{align}
	[\hat b^{(s)}_{\mathbf k},\hat b^{(s')\dagger}_{-\mathbf k'}] =\delta^{ss'} (2\pi)^3 \delta(\mathbf k + \mathbf k').
\end{align}
The polarization tensor is normalized as Eq.~\eqref{def:nom:pol}.
Then, the commutator of the scalar field is evaluated with Eq.~\eqref{md_u:rd}, and the Green function is recast into
\begin{align}
	G(x,y) & =  \frac{a(y^0)}{a(x^0)} \theta(x^0-y^0) 
	\notag 
	\\
	&\times   \int \frac{d^3 k }{(2\pi)^3} e^{i\mathbf k \cdot (\mathbf x - \mathbf y)}
	\frac{\sin (k(x^0-y^0))}{k}.\label{green:scal}
\end{align}
Similarly, the tensor Green function is found as
\begin{align}
	&G_{ij}{}^{kl}(x,y)= \frac{a(y^0)}{a(x^0)} \theta(x^0-y^0)   
	\notag 
	\\
	&\times \int \frac{d^3 k }{(2\pi)^3} e^{i\mathbf k \cdot (\mathbf x - \mathbf y)}\Pi(\mathbf k)_{ij}{}^{kl}
	\frac{\sin (k(x^0-y^0))}{k},
	\label{green:tens}
\end{align}
where the transverse-traceless projection operator is defined as
\begin{align}
	\Pi(\mathbf k)^{kl}{}_{ij} \equiv \sum_{ss'}\delta_{ss'} e^{(s)kl}(\mathbf k) e^{(s')*}_{ij}(\mathbf k) .\label{polpro}
\end{align}
Using the polarization tensor, the scalar, and tensor Green functions are related to
\begin{align}
	G_{ij}{}^{kl}(x,y) =\Pi(-i\partial)_{ij}{}^{kl} G(x,y).
\end{align}
The retarded Green function is uniquely determined for a given equation of motion, independent from a quantum state.
One can take various limits for the mode functions to simplify the formulas, but one needs to take limits such that the Wronskian condition is satisfied to find the retarded Green functions consistently. For example, the Wronskian condition is not satisfied for the limit in Eq.~\eqref{udsdrlim} because the decaying mode is dropped.
On the other hand, the Keldysh Green functions depend on a quantum state by definition. 
For a thermal state of the inverse temperature $\beta$, we find
\begin{align}
	&{\rm Tr}\left[ \hat \varrho_\chi ( \chi_{\mathbf k_2}(y^0)\chi_{\mathbf k_3}(z^0) + \chi_{\mathbf k_3}(z^0) \chi_{\mathbf k_2}(y^0) )\right] 
	\notag 
	\\
	&=
	 \left[ u^{\rm RD}_{ k_2}(y^0) u^{\rm RD*}_{ k_3}(z^0)  + u^{\rm RD*}_{ k_2}(y^0) u^{\rm RD}_{ k_3}(z^0) 
	 \right]
	 \notag 
	\\
	&
	\times (1 + 2 f_{\beta k_2}) (2\pi)^3\delta(\mathbf k_2 + \mathbf k_3).
\end{align}
Then, in the high frequency limit, $k_2\to \infty$, we find
\begin{align}
	G^K(y,z) &\simeq \int \frac{d^3k_2}{(2\pi)^3}e^{i\mathbf k_2 \cdot (\mathbf y-\mathbf z) }  \frac{\cos [k_2(y^0-z^0)]f_{\beta k_2}}{k_2 a(y^0)a(z^0)},\label{GKexp}
\end{align}
where we dropped the zero temperature part, which is renormalized to zero in the standard loop analysis.
The high-frequency limit of the integrand implies that we take the high-temperature limit after integration.

\section{Polarization tensor}
\label{propol}
Symmetry in the indices writes Eq.~\eqref{polpro} as
\begin{align}
	\Pi(\mathbf k)^{kl}{}_{ij} &=A(\delta^k{}_i\delta^l{}_j + \delta^l{}_i\delta^k{}_j) - B\delta_{ij}\delta^{kl} 
	\notag 
	\\
	&+ C( \delta_{ij}\hat k^k\hat k^l + \delta^{kl}\hat k_i\hat k_j)
	\notag 
	\\
	& - D ( \delta^{k}{}_{i}\hat k_j\hat k^l +  \delta^{l}{}_{i}\hat k_j\hat k^k +  \delta^k{}_{j}\hat k_i\hat k^l + \delta^l{}_{j}\hat k_i\hat k^k )
	\notag 
	\\
	&+E \hat k_i\hat k_j \hat k^k \hat k^l.
\end{align}
Transverse condition $\hat k_k\Pi^{kl}{}_{ij} = 0$ leads to
\begin{align}
	A-D = 0,~C-B=0,C-2 D+E = 0.
\end{align}
Similarly, the traceless condition $\delta_{kl}\Pi^{kl}{}_{ij} = 0$ turns into
\begin{align}
	2A - 3B + C= 0,~3C - 4D +E=0.
\end{align}
There are four independent equations for five variables. Hence, the transverse-traceless condition fixes the coefficients up to the normalization factor.
The normalization factor is given by imposing Eq.~\eqref{def:nom:pol}.
Then we find
\begin{align}
	A=B=C=D=E=\frac{1}{2}.
\end{align}
Hence the projection operator is expanded into
\begin{align}
	2\Pi(\mathbf k)^{kl}{}_{ij} &=\delta^k{}_i\delta^l{}_j + \delta^l{}_i\delta^k{}_j - \delta_{ij}\delta^{kl} 
	\notag 
	\\
	&
	+ \delta_{ij}\hat k^k\hat k^l + \delta^{kl}\hat k_i\hat k_j 
	- \delta^{k}{}_{i}\hat k_j\hat k^l- \delta^{l}{}_{i}\hat k_j\hat k^k
	\notag 
	\\
	&
	- \delta^k{}_{j}\hat k_i\hat k^l
	- \delta^l{}_{j}\hat k_i\hat k^k + \hat k_i\hat k_j \hat k^k \hat k^l.\label{Piexpand}
\end{align}
From Eq.~\eqref{Piexpand}, the evaluation of the angular factor is straightforward.
We find
\begin{align}
	&2\Pi(\mathbf q)^{klij}k_{1k}k_{2l}k_{1i}k_{2j} 
	\notag 
	\\
	&= k_1^2 k_2^2 \left[ 1- (\hat k_1 \cdot \hat q)^2 \right]\left[ 1-  (\hat k_2 \cdot \hat q)^2 \right],
\end{align}
and then the momentum conservation $\mathbf k_1+\mathbf k_2=\mathbf q$ and the Heron's formula yield
\begin{align}
	&\Pi(\mathbf q)^{klij}k_{1k}k_{2l}k_{1i}k_{2j} 
	\notag
	\\
	&=\frac{1}{32q^4}(k_1+k_2+q)^2(k_1+k_2-q)^2
	\notag 
	\\
	&\times (k_1-k_2+q)^2(-k_1+k_2+q)^2.\label{C7}
\end{align}

\bibliography{sample.bib}{}
\bibliographystyle{unsrturl}

\end{document}